\documentstyle[12pt,hyperref]{article}
\def\ZZZ{{\hbox{ Z\kern-1.6mm Z}}}
\def\zzz{{\hbox{ z\kern-1mm z}}}

\newcommand{\vt}{\vartheta}

\newcommand{\CC}{{\cal C}}
\newcommand{\DD}{{\cal D}}

\newcommand{\OO}{{\cal O}}

\newcommand{\wt}{\widetilde}
\newcommand{\wh}{\widehat}

\newcommand{\NN}{{\cal N}}

\newcommand{\be}{\begin{equation}}
\newcommand{\ee}{\end{equation}}
\newcommand{\ben}{\begin{eqnarray}\displaystyle}
\newcommand{\een}{\end{eqnarray}}
\newcommand{\refb}[1]{(\ref{#1})}
\newcommand{\p}{\partial}
\newcommand{\sectiono}[1]{\section{#1}\setcounter{equation}{0}}

\def\one{{\hbox{ 1\kern-.8mm l}}}
\def\zero{{\hbox{ 0\kern-1.5mm 0}}}

\begin{document}
{}~
{}~

\hfill\vbox{\hbox{hep-th/0510147}
}\break

\vskip .6cm

{\baselineskip20pt
\begin{center}
{\Large \bf
Dyon Spectrum in CHL Models
}

\end{center} }

\vskip .6cm
\medskip

\vspace*{4.0ex}

\centerline{\large \rm Dileep P. Jatkar and Ashoke Sen}

\vspace*{4.0ex}

\centerline{\large \it Harish-Chandra Research Institute}

\centerline{\large \it  Chhatnag Road, Jhusi,
Allahabad 211019, INDIA}

\vspace*{1.0ex}

\centerline{E-mail: dileep@mri.ernet.in, ashoke.sen@cern.ch,
sen@mri.ernet.in}

\vspace*{5.0ex}

\centerline{\bf Abstract} \bigskip

We propose a formula for the degeneracy of quarter BPS dyons in a
class of CHL models. The formula uses a modular form of a subgroup
of the genus two modular group
$Sp(2,\ZZZ)$. Our proposal is S-duality invariant and reproduces
correctly the entropy of a dyonic black hole to first non-leading
order for large values of the charges.

\vfill \eject

\baselineskip=18pt

\tableofcontents

\sectiono{Introduction and Summary} \label{s1}

Some years ago, Dijkgraaf, Verlinde and Verlinde\cite{9607026}
proposed a formula for the exact degeneracy of dyons in toroidally
compactified heterotic string theory.\footnote{Throughout this paper
  we shall use the word degeneracy to denote an appropriate helicity
  trace that vanishes for long supermultiplets but is non-zero for the
  intermediate supermultiplets which describe quarter BPS dyons. In
  simple terms this corresponds to counting supermultiplets with
  weight $\pm 1$, the weight being $+1$ ($-1$) if the lowest helicity
  state in the supermultiplet is bosonic (fermionic).}  By now there
has been substantial progress towards verifying this
formula\cite{0412287,0505094,0506249,0508174} and also extending it to
toroidally compactified type II string theory\cite{0506151}. These
formul\ae\ are invariant under the S-duality transformations of the
theory, and also reproduce the entropy of a dyonic black hole in the
limit of large charges\cite{0412287}.

In this paper we generalize this proposal to a class of CHL
models\cite{CHL,CP,9507027,9507050,9508144,9508154}.  The construction
of these models begins with heterotic string theory compactified on a
six torus $T^4\times S^1\times \wt S^1$, and modding out the theory by
a $\ZZZ_N$ transformation that involves $1/N$ unit of translation
along $\wt S^1$ and a non-trivial action on the internal conformal
field theory (CFT) describing heterotic string compactification on
$T^4$.  Using string-string
duality\cite{9410167,9501030,9503124,9504027,9504047} one can relate
these models to $\ZZZ_N$ orbifolds of type IIA string theory on
$K3\times S^1\times \wt S^1$, with the $\ZZZ_N$ transformation acting
as $1/N$ unit of shift along $\wt S^1$ together with an action on the
internal CFT describing type IIA string compactification on $K3$.
Under this map the S-duality of the heterotic string theory gets
related to T-duality of the type IIA string theory, which in turn can be
determined by analyzing the symmetries of the corresponding conformal
field theory. Using this procedure one finds that the orbifolding
procedure breaks the S-duality group of the heterotic string theory
from $SL(2,Z)$ to a subgroup
$\Gamma_1(N)$ of $SL(2,Z)$\cite{0502126}. This acts on the
electric charge vector $Q_e$ and the magnetic charge vector $Q_m$
as
\ben \label{eisd3}
   Q_e\to a Q_e + b Q_m, \qquad Q_m\to c Q_e + d Q_m\, ,
   \nonumber \\
   a,b,c,d\in \ZZZ, \quad
   ad-bc=1, \quad \hbox{$c=0$ mod $N$, \quad
     $a,d=1$ mod $N$} \, .
\een

The spectra of these models contain 1/4 BPS dyons.  We consider a
class of dyon states in this theory, and propose a formula for the
degeneracy of these dyons.  For technical reasons we have to restrict
our analysis to the case of prime values of $N$ only, -- among known
models the allowed prime values of $N$ are 1, 2, 3, 5 and 7.  Since
the analysis is somewhat technical, we shall summarize the proposal
here.
\begin{itemize}
\item We first define a set of coefficients $f^{(k)}_n$ ($n\ge 1$) through
the relations:
\be \label{eftau0}
f^{(k)}(\tau)\equiv \eta(\tau)^{k+2}\, \eta(N\tau)^{k+2}\, ,
\ee
\be \label{ei1}
 \sum_{n\ge 1} f^{(k)}_n e^{2\pi i \tau
     (n-{1\over 4})} = \eta(\tau)^{-6}\, f^{(k)}(\tau) \, ,
\ee
where $\eta(\tau)$ is the Dedekind  function and
\be \label{ei2}
   k = {24\over N+1} -2\, .
\ee
$f^{(k)}(\tau)$ is the unique cusp form of weight $(k+2)$ of
the S-duality group $\Gamma_1(N)$ described in \refb{eisd3}.

\item Next we define the coefficients $C(m)$ through 
\ben
\label{eisk2} C(m) = (-1)^m 
\sum_{s,n\in\zzz \atop n\ge 1} f^{(k)}_n
\delta_{4n+{s^2-1 },{m }}\, . 
\een 
As will be explained in appendix \ref{sb}, the $C(m)$'s are related to
the Fourier coefficients of a (weak) Jacobi form of the group
$\Gamma_1(N)$.

\item We now define
\be \label{eisk3}
   \Phi_k(\rho,\sigma, v) = \sum_{n,m,r\in\zzz\atop
     n,m\ge 1, \, r^2< 4mn}\, a(n,m,r)
   \, e^{2\pi i (n\rho+m\sigma+rv)}
   \, ,
\ee
where
\be \label{eisk4}
   a(n, m,r) = \sum_{\alpha\in \zzz;\alpha>0\atop
     \alpha|(n,m,r), \, {\small \rm g.c.d.}
     (\alpha,N)=1} \, \chi(\alpha)\,
   \alpha^{k-1}\, C\left({4mn
       -r^2\over \alpha^2}\right)\, ,
\ee
and $\chi(\alpha)$ takes values $1$ or $-1$ depending on the values of
$N$ and of $\alpha$ mod $N$. As explained in appendix \ref{sb},
$\chi(a)$ is a Dirichlet character mod 2 of the modular transformation
$\pmatrix{a & b\cr c & d}\in \Gamma_0(N)$, \i.e.\ it describes a
homomorphism map from $\Gamma_0(N)$ to $\ZZZ_2$.  If $k$ defined in
\refb{ei2} is even (\i.e.\ if $N=1,2,3$ or 5) then $\chi(\alpha)=1$
identically. For $N=7$ we have
\ben \label{echivalue}
   \chi(\alpha) &=& 1 \quad \hbox{for} \quad
   \alpha=1,2,4 \, \hbox{mod} \, 7\, , \nonumber \\
   &=& -1 \quad \hbox{for} \quad
   \alpha=3,5,6 \, \hbox{mod} \, 7\, .
\een 
It has been shown in appendix \ref{sa} that $\Phi_k$, defined in
\refb{eisk3}, transforms as a modular form of weight $k$ under a
subgroup $G$ of $Sp(2,\ZZZ)$ defined in \refb{egroup}.

\item In the next step we define:
\be \label{ei2.4}
   \wt\Phi_k(\wt\rho, \wt\sigma, \wt v)
   = \wt\sigma^{-k}
   \Phi_k\left(\wt\rho -{\wt v^2\over \wt\sigma}, -{1\over \wt\sigma},
     {\wt v\over \wt\sigma}\right)\, .
\ee
 $\wt\Phi_k$ transforms as a modular form of weight $k$ under a
{\it different subgroup} $\wt G$ of $Sp(2,\ZZZ)$ defined in
\refb{e2.1}, \refb{e2.2}.

\item Now consider a dyon carrying electric charge vector $Q_e$ and
  magnetic charge vector $Q_m$.  The precise conventions for defining
  $Q_e$ and $Q_m$ and their inner products have been described in
  eqs.\refb{e02}, \refb{e02a}.  We consider those dyons whose electric
  charge arises from a twisted sector of the theory, carrying
  fractional winding number along $\wt S^1$.  According to our
  proposal the degeneracy $d(Q)$ of such a dyon is given by:
\be \label{ei44.1}
   d(Q)=g\left({1\over 2}Q_m^2, {1\over 2}
     Q_e^2, Q_e\cdot Q_m\right)\, ,
\ee
where $g(m,n,p)$ is defined through the Fourier expansion
\be \label{ei44.2}
   {1\over \wt\Phi_k(\wt\Omega)} ={1\over N\, K}
   \sum_{m,n,p\atop m\ge -1,n\ge -1/N}
   e^{2i\pi(m\wt\rho
     + n\wt\sigma + p\wt v)} g(m,n,p)\, ,
\ee
and $K$ is an appropriate normalization factor given in
eqs.\refb{ecvalue1}, \refb{enknorm}.  We shall see that in
\refb{ei44.2} the sum over $m$ and $p$ run over integer values,
whereas the sum over $n$ runs over integer multiples of $1/N$.
\end{itemize}
In section \ref{s2} we explain in detail the construction outlined
above. The proposal made in \cite{9607026} is a special case ($N=1$,
$k=10$) of our more general proposal.

In section \ref{s3} we subject our proposal to some consistency tests.
First we prove, using the modular transformation properties of
$\wt\Phi_k$, that the proposed formula for $d(Q)$ is invariant under
the S-duality transformation \refb{eisd3}.  Next we prove, by studying
the explicit form of the Fourier expansion of $\wt\Phi_k$, that the
degeneracies $d(Q)$ defined through \refb{ei44.1}, \refb{ei44.2} are
integers.  Finally we analyze the behaviour of the proposed formula in
the limit of large $Q_e^2$, $Q_m^2$ and $Q_e\cdot Q_m$. Defining the
statistical entropy $S_{stat}$ as $\ln d(Q)$, we find that for large
charges it is given by:
\ben \label{ei45.8}
S_{stat} &=& {\pi\over 2} \bigg[{a^2 +S^2\over S} Q_m^2 +
   {1\over S} Q_e^2 - 2 \, {a\over S}\,  Q_e\cdot Q_m 
   +128\, \pi \, \phi(a,S) \bigg]\nonumber \\
   && + \hbox{constant} + \OO(Q^{-2})\, ,
\een
\be \label{eiphias} \phi(a,S) \equiv -{1\over 64\pi^2} \, \bigg\{
(k+2)\, \ln S +\ln f^{(k)}(a+iS) + \ln f^{(k)}(a+iS)^* \bigg\} 
\ee
where $f^{(k)}(\tau)$ has been defined in eq.\refb{eftau0} and the
variables $S$ and $a$ are to be determined by extremizing
\refb{ei45.8}.  The entropy of a black hole in these theories carrying
the same charges was computed in \cite{0508042}, generalizing earlier
results of \cite{9906094,0007195} for toroidal compactification.
\refb{ei45.8} agrees with this formula for large charges not only at
the leading order but also at the first non-leading order. This
generalizes a similar result found in \cite{0412287} for toroidal
compactification.

Appendices \ref{sb} and \ref{sa} are devoted to proving that $\Phi_k$
defined in \refb{eisk3} transforms as a modular form under an
appropriate subgroup of $Sp(2,\ZZZ)$. Modular transformation
properties of $\wt\Phi_k$ follow from these results. Appendix \ref{sc}
is devoted to proving that up to an overall normalization factor,
$\wt\Phi_k$ has a Fourier expansion in $\wt\rho$, $\wt \sigma$ and
$\wt v$ with integer coefficients, and the leading term with unit
coefficient.  This is necessary for proving that $g(m,n,p)$ defined
through \refb{ei44.2} are integers.  We also give an algorithm for
computing these coefficients.

Before concluding this section we must mention a possible caveat in
our proposal. According to \refb{ei44.1}, \refb{ei44.2} the result for
$d(Q)$ depends only on the combinations $Q_e^2$, $Q_m^2$ and $Q_e\cdot
Q_m$ and not on the details of the charge vectors.  This would be true
had all the charge vectors with given $Q_e^2$, $Q_m^2$ and $Q_e\cdot
Q_m$ been related by appropriate T-duality transformations.  However
since the same $Q_e^2$ value may arise from both the twisted and the
untwisted sector of the orbifold model which are not related by a
T-duality transformation, it is not {\it a priori} obvious that the
formula for the degeneracy should nevertheless depend only on the
combinations $Q_e^2$, $Q_m^2$ and $Q_e\cdot Q_m$.  We have already
pointed out that the proposed formula is suitable only for states
carrying twisted sector electric charges.  It is conceivable that
there are further restrictions on the charge vector $Q$ for which our
proposal for $d(Q)$ holds. In the absence of a derivation of
\refb{ei44.1}, \refb{ei44.2} from first principles ({\it e.g.} along
the lines of \cite{0505094}) we are unable to address this issue in
more detail.

\sectiono{Proposal for the dyon spectrum} \label{s2}

In this section we shall first give a brief introduction to the class
of CHL models which we shall study, then introduce the necessary
mathematical background involving modular forms of appropriate
subgroups of $Sp(2,\ZZZ)$, and finally write down our proposal for the
degeneracy of dyons in the CHL model in terms of these modular forms.

\subsection{CHL models} \label{s2.1}

We consider a class of CHL string
compactification\cite{CHL,CP,9507027,9507050,9508144,9508154} where we
begin with a heterotic string theory compactified on a six torus
$T^4\times \wt S^1\times S^1$ and mod out the theory by a $\ZZZ_N$
transformation that acts as a $1/N$ unit of shift along $\wt S^1$
together with some $\ZZZ_N$ action on the internal conformal field
theory associated with $T^4$ and the 16 left-moving chiral bosons of
heterotic string theory.  There is a dual description of the theory as
a $\ZZZ_N$ orbifold of type IIA string theory on $K3\times \wt
S^1\times S^1$ where the $\ZZZ_N$ acts as a $1/N$ unit of translation
along $\wt S^1$, together with some action on the internal conformal
field theory associated with the $K3$ compactification.  We shall
restrict our analysis to the case of prime values of $N$ since some of
the uniqueness results which we shall use hold only when $N$ is prime.
In this case the rank of the gauge group is reduced to
\be \label{e00}
   r = {48\over N+1}+ 4\,
\ee
from its original value of 28 for toroidal
compactification\cite{9508144,9508154}.  We also define:
\be \label{e01}
   k = {r-4\over 2} - 2 = {24\over N+1} - 2\, .
\ee
For $N$ prime we have the following values of $k$: $(N,k) = (2,6), (3,
4), (5,2), (7, {1})$.\footnote{Although there is no known CHL model
  with $N=11$, our analysis goes through for $N=11$ as well. This
  leads us to suspect that there may be a consistent CHL model with
  $N=11$. In this case $k=0$ and $r=8$. These eight gauge fields
  include all the six right-handed gauge fields (as is required by
  $\NN=4$ supersymmety) and the two left handed gauge fields
  associated with $\wt S^1\times S^1$. Thus all the left-handed gauge
  fields associated with $T^4$ are projected out. In the dual type IIA
  construction the orifolding must project out all the 19 anti-self
  dual two forms on $K3$ as well as a linear combination of the zero
  form and the four form. The latter result suggests that the orbifold
  action cannot be geometric, and must form part of the mirror
  symmetry group of $K3$.}  The case of toroidal compactification
corresponds to $(N,k)=(1,10)$.

We shall analyze the degeneracy of dyons carrying momentum and winding
charges as well as Kaluza-Klein and $H$-monopole charges\footnote{An
  H-monopole associated with $\wt S^1$ corresponds to a 5-brane
  transverse to $\wt S^1$ and the three non-compact directions.}
along various $\ZZZ_N$ invariant compact directions.  Let $\wt n$,
$\wt w$, $\wt K$ and $\wt H$ denote the number of units of momentum,
winding, Kaluza-Klein monopole and H-monopole charges associated with
the circle $\wt S^1$.  Then we define
\be \label{e02}
   Q_e = \pmatrix{\wt n \cr \wt w \cr \vdots  }, \qquad
   Q_m=\pmatrix{\wt H\cr \wt K\cr \vdots  },  \qquad
   Q=\pmatrix{   Q_m\cr    Q_e}\, ,
\ee
where $\cdots$ in the expression for $Q_e$ denote the momentum and
winding along $S^1$ and the other $\ZZZ_N$ invariant directions 
associated with the internal conformal field theory
and $\cdots$ in the expression for $Q_m$ denote their
magnetic counterpart. We also define
\be \label{e02a}
   Q_e^2 = 2( \wt n\wt w+\cdots), \quad    Q_m^2=2
   ( \wt K\wt H+\cdots), \quad
   Q_e\cdot    Q_m = ( \wt n \wt K + \wt w \wt H+\cdots)\, ,
\ee
where $\cdots$ denote the contribution from the other components of
the charges. Now note that while the charge quantum numbers $\wt n$
and $\wt K$ are integers, the winding charge $\wt w$ associated with
the circle $\wt S^1$ can take values in units of $1/N$ once we include
twisted sector states in the spectrum. Also the H-monopole charge $\wt
H$ associated with $\wt S^1$ is quantized in units of $N$, since in
order to get a $\ZZZ_N$ invariant configuration of five-branes
transverse to $\wt S^1$ we need to have $N$ equispaced five-branes on
$\wt S^1$. The contribution from $\cdots$ terms are quantized in
integer units.  As a result, $Q_e^2/2$ is quantized in units of $1/N$,
whereas $Q_m^2/2$ and $Q_e\cdot Q_m$ are quantized in integer units.

The S-duality symmetry of the theory is generated by a set of $2\times
2$ matrices of the form\cite{9507050,9508154,0502126}:
\be \label{esd1}
   \pmatrix{a & b\cr c & d}, \qquad a,b,c,d\in \ZZZ, \quad
   ad-bc=1, \quad \hbox{$c=0$ mod $N$, \quad
     $a,d=1$ mod $N$} \, .
\ee
This group of matrices is known as $\Gamma_1(N)$.  The S-duality
transformation acts on the electric and the magnetic charge vectors
as:
\be \label{esd3}
   Q_e\to a Q_e + b Q_m, \qquad Q_m\to c Q_e + d Q_m\, ,
   \qquad \pmatrix{a & b\cr c & d} \in \Gamma_1(N)\, .
\ee
Our goal is to propose a formula for the degeneracy $d(Q)$ of dyons in
this theory carrying charge vector $Q$ such that it is invariant under
the S-duality transformation \refb{esd3}, and also, for large charges,
$\ln d(Q)$ reproduces the entropy of the BPS black hole carrying
charge $Q$.

\subsection{The modular form $\Phi_k$} \label{s2.2}

Since the proposed expression for $d(Q)$ involves an integration over
the period matrices of the genus two Riemann surfaces, we need to
begin by reviewing some facts about this period matrix. It is
parametrized by the $2\times 2$ complex symmetric matrix
\be \label{e1}
   \Omega = \pmatrix{\rho & v \cr v & \sigma} \, ,
\ee
subject to the restriction   
\be \label{elight1} Im(\rho)>0, \qquad Im(\sigma)>0, \qquad (Im \,
\rho) \, (Im \, \sigma) > (Im\, v)^2\, . \ee The modular group
$Sp(2,\ZZZ)$ transformation acts on this as \be \label{e2}
   \Omega \to \Omega'=(A\Omega +B) (C\Omega +D)^{-1}\, ,
\ee
where $A,B,C,D$ are $2\times 2$ matrices satisfying
\be \label{e3}
   AB^T=BA^T, \qquad  CD^T=DC^T, \qquad AD^T-BC^T=I\, .
\ee Let us denote by $G$ the subgroup of $Sp(2,\ZZZ)$ generated by
the $4\times 4$ matrices
\ben \label{egroup}
   g_1(a,b,c,d) &\equiv& \pmatrix{ a & 0 & b & 0 \cr
     0 & 1 & 0 & 0\cr c & 0 & d & 0\cr 0 & 0 & 0 & 1}\, ,
   \qquad ad-bc=1, \quad \hbox{$c=0$ mod $N$, \quad $a,d=1$
     mod $N$}
   \nonumber \\
   g_2 &\equiv& \pmatrix{0 & 1 & 0 & 0 \cr -1 & 0 & 0 & 0\cr
     0 & 0 & 0 & 1\cr 0 & 0 & -1 & 0}\, , \nonumber \\
   g_3(\lambda, \mu) &\equiv& \pmatrix{ 1 & 0 & 0 & \mu \cr
     \lambda & 1 & \mu & 0\cr 0 & 0 & 1 & -\lambda\cr
     0 & 0 & 0 & 1}\, , \qquad \lambda, \mu \in \ZZZ.
   \nonumber \\
\een
{}From the definition it follows that
\be \label{egene1}
   \pmatrix{A & B\cr C & D}\in G \quad
   \rightarrow \quad \hbox{$C={\bf 0}$
     mod $N$, \quad $\det A = 1$
     mod $N$, \quad $\det D=1$ mod $N$
   }
   \, .
\ee 
We suspect that the reverse is also true, {\i.e.} if $C= {\bf 0}$ mod
$N$ and $\det A=1$ mod $N$ (which implies that $\det D=1$ mod $N$)
then the $Sp(2,\ZZZ)$ matrix $\pmatrix{A & B\cr C & D}$ belongs to
$G$. However at present we do not have a proof of this, and we shall
not use this result in our analysis.

The group $G$ has the following properties:
\begin{enumerate}
\item It contains all elements of the form:
\ben \label{e4}
   && g_1(a,b,c,d) g_2 g_1(a,-b,-c,d)  (g_2)^{-1}
   = \pmatrix{a
     & 0 & b & 0\cr 0 & a & 0 & -b\cr
     c & 0 & d & 0\cr 0 & -c & 0 & d}
   \, ,
   \nonumber \\
   &&
   \qquad
   ad-bc=1, \qquad 
   c =  \hbox{0 mod $N$}, \qquad a,d=\hbox{1 mod $N$}\, .
\een
We shall denote by $H$ the subgroup of $G$ containing elements of the
form \refb{e4}. It is isomorphic to $\Gamma_1(N)$ introduced in
\refb{esd1}.

\item As shown in appendix \ref{sa}, there is a modular
  form\footnote{Throughout this paper we shall refer to as modular
    forms the functions of $\rho$, $\sigma$, $v$ which transform as
    \refb{e5} under a subgroup of $Sp(2,\ZZZ)$ but which may have
    poles in the Siegel upper half plane or at the cusps.  Thus these
    modular forms are not necessarily entire.}  $\Phi_k(\Omega)$ of
  $G$ of weight $k$ obeying the usual relations
\be \label{e5}
   \Phi_k\left((A\Omega +B) (C\Omega
   +D)^{-1}\right) = \{ \det(C\Omega +D)\}^k \,
   \Phi_k(\Omega)
    \, , \qquad \pmatrix{A & B\cr C & D} \in G\, .
\ee 
In fact as has been shown in eqs.\refb{eyy1}, \refb{eyy2}, the
transformation law \refb{e5} holds also for the $Sp(2,\ZZZ)$ element
\be \label{eyy3}
\pmatrix{A & B\cr C & D}=\pmatrix{a
     & 0 & b & 0\cr 0 & a & 0 & -b\cr
     c & 0 & d & 0\cr 0 & -c & 0 & d}, \qquad
     \pmatrix{a & b \cr c & d} \in \Gamma_0(N)
   \, ,
\ee
where $\Gamma_0(N)$ is defined as the collection of $2\times 2$
matrices of the form
\be \label{esd2}
   \pmatrix{a & b\cr c & d}, \qquad a,b,c,d\in \ZZZ, \quad
   ad-bc=1, \quad \hbox{$c=0$ mod $N$ } \, .
\ee

It has also been shown in appendix \ref{sa} that as $v\to 0$,
\be \label{e6}
   \Phi_k(\Omega) \simeq 4\pi^2\,
   v^2 \,  f^{(k)}(\rho) \, f^{(k)}(\sigma) + \OO(v^4)\, ,
\ee
where
\be \label{emod}
   f^{(k)}(\tau) = (\eta(\tau))^{k+2} \, (\eta(N\tau))^{k+2}\,
\ee
and $\eta(\tau)$ is the Dedekind $\eta$ function.
$f^{(k)}(\tau)$ is the unique cusp form
of $\Gamma_1(N)$ of weight $k+2=24/(N+1)$\cite{mod}:
\ben \label{fcusp1}
&&   f^{(k)}\left( {a\tau + b\over c\tau +d}\right) 
= (c\tau + d)^{k+2}
   f^{(k)}(\tau)\, , \nonumber \\
&& \lim_{\tau\to i\infty} f^{(k)}(\tau)=0\, ,
   \quad \lim_{\tau \to p/q} f^{(k)}(\tau) = 0\, \quad
   \hbox{for} \quad p,q\in \ZZZ\, .
\een

An algorithm for constructing the modular form $\Phi_k$ in terms of
the cusp forms $f^{(k)}(\tau)$ of $\Gamma_1(N)$ can be found along the
lines of the analysis performed in \cite{eichler,skor,rama} and has
been described in appendices \ref{sb} and \ref{sa}.  The results have
already been summarized in eqs.\refb{ei1}-\refb{eisk4}.
\end{enumerate}

Neither the group $G$ nor the modular form $\Phi_k$ will be used
directly in writing down our proposal for $d(Q)$. Instead we shall
use them to define a modular form $\wt\Phi_k$ of a different
subgroup $\wt G$ of $Sp(2,\ZZZ)$ which is related to $G$ by a
conjugation. It will be this modular form $\wt\Phi_k$ that will be
used in writing down our proposal.

\subsection{The modular form $\wt\Phi_k$}

We now introduce the $Sp(2,\ZZZ)$ matrices:
\be \label{e2.1}
   U_0\equiv \pmatrix{A_0 & B_0\cr C_0 & D_0}=
   \pmatrix{1 & 0 & 0 & 0\cr 0 & 0 & 0 & 1\cr
     0 & 0 & 1 & 1\cr 1 & -1 & 0 & 0}\, , \qquad
   U_0^{-1} = \pmatrix{ 1 & 0 & 0 & 0 \cr 1 & 0
     & 0 & -1\cr 0 & -1 & 1 & 0\cr
     0 & 1 & 0 & 0}\, ,
\ee 
and denote by $\wt G$ the subgroup of $Sp(2,\ZZZ)$ containing
elements of the form:
\be \label{e2.2}
   U_0 U U_0^{-1}, \qquad U\in G\, .
\ee
Clearly $\wt G$ is isomorphic to $G$. We also define
\be \label{e2.3}
   \wt\Omega = (A_0\Omega +B_0)(C_0\Omega +D_0)^{-1}
   \equiv \pmatrix{\wt \rho & \wt v \cr \wt v & \wt\sigma} \, ,
\ee
and
\be \label{e2.4}
   \wt \Phi_k(\wt \Omega) = \{\det (C_0\Omega + D_0)\}^k
   \Phi_k(\Omega) = (2v-\rho-\sigma)^k \,
   \Phi_k(\Omega) \, .
\ee
Using \refb{e5} and \refb{e2.2}-\refb{e2.4} one can show that for
\be \label{e2.5}
   \wt\Omega'=(\wt A \wt\Omega +\wt B)(\wt C\wt \Omega
   +\wt D)^{-1} \, , \qquad \pmatrix{\wt A & \wt B\cr
     \wt C & \wt D}\in \wt G\, ,
\ee
we have
\be \label{e2.6}
   \wt \Phi_k(\wt\Omega') = \{\det (\wt C \wt \Omega +
   \wt D)\}^k \, \wt \Phi_k(\wt\Omega)\, .
\ee

The group $\wt G$ includes the translation symmetries:
\be \label{etrans}
   (\wt\rho, \wt\sigma, \wt v) \to (\wt\rho+1, \wt\sigma, \wt v),
   \qquad (\wt\rho, \wt\sigma, \wt v)\to (\wt\rho, \wt\sigma+N, \wt v),
   \qquad (\wt\rho, \wt\sigma, \wt v) \to (\wt\rho, \wt\sigma, \wt v+1)\, ,
\ee
under which $\wt \Phi_k$ is invariant. The corresponding
elements of $G$ are
\ben \label{etrsg}
   \pmatrix{1 & 0 & 1 & 1\cr 0 & 1 & 1 & 1\cr 0 & 0 & 1 & 0\cr
     0 & 0 & 0 & 1} &=& g_1(1,1,0,1) g_2 g_1(1,1,0,1)
   (g_2)^{-1} g_3(0,1)\nonumber \\
   \pmatrix{1 & 0 & 0 & 0\cr
     0 & 1 & 0 & 0\cr -N & N & 1 & 0\cr N & -N & 0 & 1}
   &=& g_2 g_3(-1,0) g_2 g_1(1,0,-N,1) g_2 g_3(1,0) g_2
   \nonumber \\
   \hbox{and} \qquad \pmatrix{2 & -1 & 0 & 0\cr
     1 & 0 & 0 & 0\cr 0 & 0 & 0 & -1\cr 0 & 0 & 1 & 2}
   &=& (g_2)^{-1} g_3(-2,0) \, ,
\een
respectively, with the $g_i$'s as defined in \refb{egroup}.

$\wt G$ also contains a subgroup $\wt H$
consisting of elements of the form:
\be \label{e2.7}
   U_0 U U_0^{-1}, \qquad U\in H\, .
\ee
Since the elements of $H$ are of the form \refb{e4}, we
see using \refb{e2.1} that the elements of $\wt H$ are of the form:
\be \label{e2.8}
   \pmatrix{a & -b & b & 0\cr -c & d & 0 & c\cr
     0 & 0 & d & c\cr 0 & 0 & b & a}\, , \qquad
   ad-bc=1,  \qquad a,d =  \hbox{ 1 mod $N$},
   \qquad c =  \hbox{0 mod $N$}\, .
\ee 
In fact, due to the result described in \refb{e5}, \refb{eyy3}
the modular transformation law \refb{e2.6} holds  for  a more
general class of $Sp(2,\ZZZ)$ elements of the form:
\be \label{eyy4}
\pmatrix{\wt A & \wt B\cr \wt C & \wt D} =
   \pmatrix{a & -b & b & 0\cr -c & d & 0 & c\cr
     0 & 0 & d & c\cr 0 & 0 & b & a}\, , \qquad
     \pmatrix{a & b\cr c & d} \in \Gamma_0(N)\, .
\ee

Using \refb{e1}, \refb{e2.1} and \refb{e2.3} we see that
the variables $(\wt\rho, \wt\sigma, \wt v)$ are related to the
variables $(\rho, \sigma, v)$ via the relations:
\be \label{e3.8}
   \rho = {\wt \rho \wt\sigma - \wt v^2\over \wt\sigma}, 
   \qquad \sigma = {\wt\rho \wt \sigma - (\wt v - 1)^2\over \wt 
   \sigma}
   \, , \qquad
   v = {\wt\rho \wt\sigma - \wt v^2 + \wt v\over \wt\sigma}\, ,
\ee
or equivalently,
\be \label{e3.8a}
   \wt\rho = {v^2-\rho\sigma \over 2v-\rho-\sigma}, \qquad
   \wt\sigma={1\over 2v-\rho-\sigma}, \qquad \wt v =
   {v-\rho \over 2v-\rho-\sigma}\, .
\ee
\refb{e2.4} now gives
\be \label{ewtphi}
   \wt\Phi_k(\wt\rho, \wt\sigma, \wt v) = \wt\sigma^{-k}
   \Phi_k\left(\wt\rho -{\wt v^2\over \wt\sigma},
     \wt\rho -{(\wt v-1)^2\over \wt\sigma}, \wt\rho -
     {\wt v^2\over \wt\sigma} + {\wt v\over \wt\sigma}\right)\, .
\ee
Using the invariance of $\Phi_k$ under the transformation $g_3(-1,0)$:
\be \label{eg2phi}
   \Phi_k(\rho,\sigma,v) = \Phi_k(\rho, \sigma+\rho-2v, v-\rho)\, ,
\ee
we can rewrite \refb{ewtphi} as
\be \label{ewtphi1}
   \wt\Phi_k(\wt\rho, \wt\sigma, \wt v)
   = \wt\sigma^{-k}
   \Phi_k\left(\wt\rho -{\wt v^2\over \wt\sigma}, -{1\over \wt\sigma},
     {\wt v\over \wt\sigma}\right)\, .
\ee

Eq.\refb{e3.8} shows that the region $v\simeq 0$ corresponds to
$\wt\rho \wt\sigma - \wt v^2 + \wt v\simeq 0$. \refb{e6} and \refb{ewtphi}
now give
\be \label{e4.1}
   \wt \Phi_k(\wt\rho,\wt\sigma,\wt v) \simeq 4\pi^2
   \, \wt \sigma^{-k-2}
   ( \wt\rho \wt\sigma - \wt v^2 + \wt v)^2 f^{(k)}(\rho) f^{(k)}(\sigma)
   + \OO\left(( \wt\rho \wt\sigma - \wt v^2 + \wt v)^4
   \right)\, .
\ee
On the other hand \refb{ewtphi1} shows that near $\wt v=0$,
\be \label{exx1}
   \wt\Phi_k(\wt\rho,\wt\sigma,\wt v) \simeq 4\pi^2\,
   \wt\sigma^{-k-2}
   \wt v^2 f^{(k)}(\wt\rho)f^{(k)}(-\wt\sigma^{-1}) +\OO(\wt v^4)\, .
\ee
Using the definition of $f^{(k)}(\tau)$ given in
\refb{emod}, and the modular transformation law of $\eta(\tau)$:
\be \label{etamod}
   \eta(-1/\tau) = (-i\tau)^{1/2} \eta(\tau)\, ,
\ee
we can rewrite \refb{exx1} as
\be \label{exx2}
   \wt\Phi_k(\wt\rho,\wt\sigma,\wt v) \simeq (i\sqrt N)^{-k-2}
   \, 4\pi^2\, \wt v^2 f^{(k)}(\wt\rho)f^{(k)}( \wt\sigma/N) +\OO(\wt v^4)\, .
\ee

\subsection{The dyon spectrum} \label{s2.4}

Our proposal for $d(Q)$ involves the modular form $\wt\Phi_k$ and
can be stated as follows:
\be \label{e44.1}
   d(Q)=g\left({1\over 2}Q_m^2, {1\over 2}
     Q_e^2, Q_e\cdot Q_m\right)\, ,
\ee
where $g(m,n,p)$ is defined through the Fourier expansion:
\be \label{e44.2}
   {1\over \wt\Phi_k(\wt\Omega)} ={1\over N\, K}
   \sum_{m,Nn,p\in \zzz\atop m\ge -1,n\ge -1/N}
   e^{2i\pi(m\wt\rho
     + n\wt\sigma + p\wt v)} g(m,n,p)\, ,
\ee
$K$ being an appropriate normalization factor.  The multiplicative
factor of $1/N$ has been included for later convenience; it could have
been absorbed into the definition of $K$.  {}From eq.\refb{etrans} we
see that the sum over $m$ and $p$ run over integer values, whereas the
sum over $n$ run over integer multiples of $1/N$. Eq.\refb{e44.1} then
indicates that $Q_m^2/2$ and $Q_e\cdot Q_m$ are quantized in integer
units and $Q_e^2/2$ is quantized in units of $1/N$. This is consistent
with the analysis given below \refb{e02a}.

Eqs.\refb{e44.1}, \refb{e44.2}
may be rewritten as
\be \label{e7}
   d(Q) = K\, \int_\CC\,  d^3 \wt \Omega \, e^{-i\pi Q^T \cdot
     \wt \Omega Q}\,
   {1 \over \wt \Phi_k(\wt\Omega)}\, ,
\ee
where the integration runs over the three cycle $\CC$ defined as
\be \label{e44.3}
   Im(\wt\rho),  \, Im(\wt\sigma),  \,
   Im(\wt v) = \hbox{fixed}, \qquad
   0\le Re(\wt\rho) \le 1, \quad 0\le Re(\wt\sigma) \le N, \quad
   0\le Re(\wt v) \le 1\, ,
\ee
 \be \label{e8}
   d^3\wt\Omega =
   d\wt\rho \, d\wt\sigma\, d\wt v\, ,
\ee
and
\be \label{eqt}
Q^T \wt\Omega Q = \wt \rho Q_m^2 + \wt\sigma Q_e^2 +
2\wt v Q_e\cdot Q_m\, .
\ee

\sectiono{Consistency checks} \label{s3}

In this section we shall subject our proposal \refb{e7} to various
consistency checks. In section \ref{s3.1} we check the S-duality
invariance of \refb{e7}. In section \ref{s3.3} we check that $d(Q)$
defined in section \ref{s2.4} are integers.  In section \ref{s3.2} we
verify that for large charges \refb{e7} reproduces the black hole
entropy to first non-leading order.

\subsection{S-duality invariance of $d(Q)$} \label{s3.1}

First we shall prove the S-duality invariance of the formula
\refb{e7}. For the CHL models considered here, the S-duality group is
$\Gamma_1(N)$ under which the electric and magnetic charges transform
to;
\be \label{e3.1}
   Q_e \to    Q_e' = a    Q_e + b    Q_m,
   \qquad Q_m\to Q_m' = c    Q_e + d    Q_m\, ,
   \qquad \pmatrix{a & b\cr c & d}\in \Gamma_1(N)\, .
\ee
Let us define
\be \label{e3.2}
   \wt\Omega'\equiv\pmatrix{\wt \rho' & \wt v'
    \cr \wt v' & \wt\sigma'}
   = (\wt A \wt\Omega + \wt B) (\wt C \wt\Omega +\wt D)^{-1},
   \qquad \pmatrix{\wt A & \wt B\cr
     \wt C & \wt D}= \pmatrix{a & -b & b & 0\cr -c & d & 0 & c\cr
     0 & 0 & d & c\cr 0 & 0 & b & a}\, \in \wt H\, .
\ee
This gives
\ben \label{e3.3}
   \wt \rho' = a^2\wt\rho + b^2\wt\sigma - 2 ab \wt v + ab\, ,
   \nonumber \\
   \wt\sigma' = c^2\wt\rho + d^2 \wt\sigma - 2cd\wt v + cd\, ,
   \nonumber \\
   \wt v' = -ac\wt\rho - bd \wt\sigma + (ad + bc)\wt v- bc\, .
\een
Using \refb{e3.1}, \refb{e3.3} and the quantization laws of
$Q_e^2$, $Q_m^2$ and $Q_e\cdot Q_m$ one can easily verify that
\be \label{e3.4}
   e^{i\pi Q^T \cdot \wt \Omega Q} = e^{i\pi 
   Q^{\prime T}\cdot \wt\Omega' Q'}
   \, ,
\ee
and
\be \label{e3.5}
   d^3\wt\Omega = d^3\wt\Omega'\, .
\ee
On the other hand, eqs.\refb{e2.6},
\refb{e3.2} give
\be \label{e3.6}
   \wt\Phi_k(\wt \Omega') = \wt\Phi_k(\wt \Omega)\, .
\ee
Finally we note that under the map \refb{e3.3} the three cycle $\CC$,
which is a three torus lying along the real $\wt\rho$, $\wt\sigma$ and
$\wt v$ axes, -- with length 1 along $\wt\rho$ and $\wt v$ axis and
length $N$ along $\wt\sigma$ axis, -- gets mapped to itself up to a
shift that can be removed with the help of the shift symmetries
\refb{etrans}.  Thus eqs.\refb{e3.4}-\refb{e3.6} allow us to express
\refb{e7} as
\be \label{e3.7}
   d(Q) = K\, \int_{\CC}
   d^3 \wt \Omega' \, e^{-i\pi Q^{\prime T} \cdot
     \wt \Omega' Q'}\,
   {1 \over \wt \Phi_k(\wt\Omega')} = d(Q')\, .
\ee
This proves invariance of $d(Q)$ under the S-duality
group $\Gamma_1(N)$.

Due to eq.\refb{eyy4} the formula for $d(Q)$ is actually invariant
under a bigger group $\Gamma_0(N)$. This indicates that the full
U-duality group of the theory may contain $\Gamma_0(N)$ as a subgroup.
The physical origin of this $\Gamma_0(N)$ is best understood in the
description of the theory as a $Z_N$ orbifold of type II string theory
on $K3\times S^1\times \wt S^1$. In this case $\Gamma_0(N)$ acts as a
T-duality transformation on $\wt S^1\times S^1$.  Hence, acting on
$1/N$ unit of shift along $\wt S^1$, represented by the vector
$\pmatrix{1/N\cr 0\cr}$, the $\Gamma_0(N)$ element $\pmatrix{a & b\cr
  c & d}$ gives a vector $\pmatrix{a/N\cr 0}$ modulo integer lattice
vectors, representing $a/N$ unit of shift along $\wt S^1$.  This would
seem to take this to a different theory where the $\ZZZ_N$ generator
acts as $a/N$ units of shift along $\wt S^1$ but has the same action
on $K3$. Of course this group also has an element that contains $1/N$
unit of shift (mod 1) along $\wt S^1$, but its action on $K3$ is a
power of the action of the original $\ZZZ_N$ generator. Thus in order
that $\Gamma_0(N)$ is a symmetry, its action must be accompanied by an
internal symmetry transformation in $K3$ that permutes the action of
different $\ZZZ_N$ elements on $K3$.

\subsection{Integrality of $d(Q)$} \label{s3.3}

In this section we shall show that $d(Q)$ defined through
\refb{e44.1}, \refb{e44.2} are integers. This is a necessary condition
for them to be interpreted as the degeneracy of dyonic states in this
theory. For this we begin with the expansion of $\wt\Phi_k(\wt\rho,
\wt\sigma, \wt v)$ described in appendix \ref{sc}:\footnote{For
  definiteness we have chosen the expansion in a form that will be
  suitable for the $Im(\wt v)<0$ region. For $Im(\wt v)>0$ we need to
  change the sign of $\wt v$ in this expansion.}
\ben \label{e33.1}
   \wt\Phi_k(\wt\rho, \wt\sigma, \wt v) &=& C\, e^{2\pi i\wt\rho
     +2\pi i \wt\sigma/N + 2\pi i \wt v} \,
   \bigg( 1 - 2e^{-2\pi i \wt v} + e^{-4\pi i \wt v} \nonumber \\
   &&
   + \sum_{q,r,s\in \zzz\atop q,r\ge 0,q+r\ge 1} \, b(q,r,s)\,
   e^{2\pi i q\wt\rho
     +2\pi i r\wt\sigma/N + 2\pi i s\wt v}\bigg)\, ,
\een
with the coefficients $b(q,r,s)$ being integers and  
\be \label{ecvalue1}
   C = -(i\sqrt N)^{-k-2}\, .
\ee
This gives
\ben \label{e33.2}
   {N \, K \over   \wt\Phi_k(\wt\rho, \wt\sigma, \wt v) }
   &=& {N \, K \over C} \, e^{-2\pi i\wt\rho
     -2\pi i \wt\sigma/N - 2\pi i \wt v} \, \nonumber \\
   && \bigg[ 1 +
   \sum_{l=1}^\infty (-1)^l \bigg( - 2
   e^{-2\pi i \wt v} + e^{-4\pi i \wt v}   \nonumber \\
   &&
   +   \sum_{q,r,s\in \zzz\atop q,r\ge 0,q+r\ge 1} \, b(q,r,s)\,
   e^{2\pi i q\wt\rho
     +2\pi i r\wt\sigma/N + 2\pi i s\wt v}\bigg)^l\bigg]\, .
   \nonumber \\
\een

We can now expand the terms inside the square bracket to get an
expansion in positive powers of $e^{2\pi i\wt\rho}$ and $e^{2\pi
  i\wt\sigma/N}$ and both positive and negative powers of $e^{2\pi
  i\wt v}$. (For $\wt\rho$ and $\wt\sigma$ independent terms inside
the square bracket the expansion has only negative powers of $e^{2\pi
  i\wt v}$.) Since $b(q,r,s)$ are integers, each term in this
expansion will also be integers. Comparing \refb{e33.2} with
\refb{e44.1}, \refb{e44.2} we see that $d(Q)$ are integers as long as
we choose $K$ such that $N\, K \over C$ is an integer.  In particular
if we choose 
\be \label{enknorm} 
K={C\over N}\, , 
\ee 
then this would correspond to counting the degeneracy in multiples of
degeneracy of states with a $Q$ for which $Q_e^2/2=-1/N$, 
$Q_m^2/2=-1$
and $Q_e\cdot Q_m=-1$.

Comparison of \refb{e33.2} and \refb{e44.2} also shows that in
order to get non-zero $d(Q)$ we must have
\be \label{erange}
   {1\over 2}Q_e^2\ge -{1\over N}, \qquad {1\over 2}
   Q_m^2\ge -1\, .
\ee
The condition ${1\over 2}Q_e^2\ge -{1\over N}$ may sound surprising at
first sight, since for ordinary electrically charged states this
condition, arising from the level matching condition, takes the form
${1\over 2}Q_e^2\ge -1$. The $-1$ in the right hand side of this
inequality is the $L_0$ eigenvalue of the left-moving ground state of
the world-sheet theory. We should note however that here we are
considering states whose electric charge arises in the twisted sector
and in applying the level matching condition we must use the ground
state energy of the twisted sector. In our example, the total number
of twisted scalar fields is given by
\be \label{etw1}
   (28-r) = 24 \, {N-1\over N+1}\, .
\ee
This corresponds to $12(N-1)/(N+1)$ complex scalars. These can be
divided into $24/(N+1)$ sets, each set containing $(N-1)/2$ complex
scalars on which the orbifold group acts as a rotation by angles
$\phi_1=2\pi/N$, $\phi_2=4\pi/N$, $\ldots$ $\phi_{(N-1)/
  2}=\pi(N-1)/N$.\footnote{For $N=2$ the counting is slightly
  different. In this case each set contains a $\ZZZ_2$ even real
  scalar and a $\ZZZ_2$ odd real scalar. The final result is the same
  as \refb{etw2}.}  The net $L_0$ eigenvalue of the twisted sector
ground state then takes the form:
\be \label{etw2}
   -1 + {24\over N+1} \, {1\over 2} \, \sum_{j=1}^{(N-1)/2}
   {j\over N} \left(1 - {j\over N}\right)  = -{1\over N}\, .
\ee
This gives the restriction
\be \label{etw3}
 {1\over 2}\,  Q_e^2\ge -{1\over N}\, .
\ee
This is consistent with \refb{erange}.

\subsection{Black hole entropy} \label{s3.2}

We shall now show that in the limit of large charges the expression
for $d(Q)$ given in \refb{e7} reproduces the extremal black hole
entropy carrying the same charges.  The analysis proceeds as in
\cite{9607026,0412287}.  First of all, following the procedure of
\cite{9607026} one can deform the integration contour $\CC$ in
\refb{e7}, picking up contribution from the residues at various poles
of the integrand. These poles occur at the zeroes of $\wt\Phi_k$,
which, according to \refb{e4.1} (or \refb{exx1}) occur on the divisor
$\wt\rho\wt\sigma-\wt v^2 +\wt v=0$ and its images under
$Sp(2,\ZZZ)$.\footnote{We are assuming that $\wt\Phi_k$ does not have
  additional zeroes other than at the images of $\wt\rho\wt\sigma-\wt
  v^2 +\wt v=0$, or that even if such zeroes are present, their
  contribution to the entropy is subdominant for large charges.}  For
large charges the dominant contribution comes from the pole at which
the exponent in \refb{e7} takes maximum value at its saddle point; the
contribution from the other poles are exponentially suppressed. The
analysis of \cite{9607026} showed that the divisor that gives dominant
contribution corresponds to
\be \label{ediv}
   \DD: \quad \wt\rho\wt\sigma-\wt v^2 +\wt v=0\, .
\ee
We now carry out the $\wt v$ integral in \refb{e7} using Cauchy's
integration formula.  Let us denote by $\wt v_\pm$ the zeroes of
$(\wt\rho \wt\sigma - \wt v^2 + \wt v)$:
\be \label{e5.1}
   \wt v_\pm = {1\over 2} \pm \Lambda(\wt\rho,\wt\sigma),
   \qquad \Lambda(\wt\rho,\wt\sigma)
   = \sqrt{{1\over 4}+\wt\rho\wt\sigma}\, .
\ee
The contour integral in $\wt v$ plane is evaluated by keeping
$\wt\rho$ and $\wt\sigma$ fixed.  The zeros $\wt v_\pm$ are distinct
for generic but fixed $\wt\rho$ and $\wt\sigma$.  However, in the
$\wt\rho$, $\wt\sigma$ and $\wt v$ space the divisor $\DD$ is a
continuous locus connecting these zeros.  Therefore
deformation of the integration contour through the divisor picks
up the contribution from the pole only once; we can choose this
pole to be at 
either $\wt v_+$ or $\wt v_-$.  We will
consider the contribution coming from $\wt v = \wt v_-$.  
We see from \refb{e7}, \refb{e4.1} that near $\wt v = \wt
v_-$ the integrand behaves as:
\ben \label{e5.2}
K\,   \exp\left(- i\pi (\wt\rho    Q_m^2 + \wt\sigma    Q_e^2
     + 2\wt  v    Q_m\cdot    Q_e)\right) (4\pi^2)^{-1}
   \wt \sigma^{k+2}
   (\wt v - \wt v_+)^{-2} (\wt v - \wt v_-)^{-2}
   f^{(k)}(\rho)^{-1} f^{(k)}(\sigma)^{-1} \nonumber \\
   +\OO\left((\wt v - \wt v_-)^0\right) \, . \qquad \qquad
   \qquad
\een
Using the results
\be \label{e5.3}
   {\p \rho \over \p \wt v} = -{2 \wt v\over \wt\sigma}, \qquad
   {\p \sigma \over \p \wt v} = -{2 (\wt v-1)\over \wt\sigma},
\ee
which follow from \refb{e3.8}, we get the result of the contour
integration around $\wt v_-$ to be
\be \label{e5.4}
   d(Q)=(-1)^{Q_e\cdot Q_m}\,  (2\pi)^{-1} i \, K\,
   \int d\wt \rho d\wt\sigma \,
   e^{i\pi X(\wt\rho, \wt\sigma) + \ln \Delta(\wt \rho, \wt\sigma) }\, ,
\ee
where
\be \label{e45.1}
   X(\wt\rho,\wt\sigma)  = -\wt\rho Q_m^2 - \wt\sigma
   Q_e^2 + 2\Lambda(\wt\rho,\wt\sigma) \, Q_e\cdot Q_m
   +{k+2\over i\pi} \ln\wt\sigma - {1\over i\pi}
   \ln f^{(k)}(\rho) -  {1\over i\pi}
   \ln f^{(k)}(\sigma)\, ,
\ee
\be \label{e45.2}
   \Delta(\wt\rho,\wt\sigma) = {1\over 4
     \Lambda(\wt\rho,\wt\sigma)^2}
   \left[ -2\pi i Q_e\cdot Q_m +{1\over
       \Lambda(\wt\rho,\wt\sigma)} +
     {2\over \wt\sigma} \left\{ \wt v_- 
     {f^{(k)\prime}(\rho)\over f^{(k)}(\rho)}
       - \wt v_+ {f^{(k)\prime}
       (\sigma)\over f^{(k)}(\sigma)}\right\}\right]\, .
\ee 
In eqs.\refb{e45.1}, \refb{e45.2} $\rho$ and $\sigma$ are to be
regarded as functions of $\wt\rho$ and $\wt\sigma$ via
eq.\refb{e3.8a} at $v=0$:
 \be \label{e45.6}
   \wt \rho ={\rho\sigma\over \rho+\sigma}, \qquad
   \wt\sigma =-{1\over \rho+\sigma}\, .
\ee

We shall now use eqs.\refb{e5.4}-\refb{e45.2} to
calculate the statistical entropy
\be \label{estat}
S_{stat}=\ln |d(Q)|\, ,
\ee
for large values of the charges.  For this we shall carry out the
integration over $\wt\rho$ and $\wt\sigma$ using a saddle point
approximation, keeping terms in the entropy to leading order as well
as first non-leading order in the charges. As we shall see, at the
saddle point $\wt\rho$ and $\wt\sigma$ take finite values. Thus the
leading order contribution to the entropy comes from the first three
terms in $X$ quadratic in the charges. In order to evaluate the first
non-leading correction, we need to evaluate the order $Q^0$
contribution from the other terms at the saddle point; but
determination of the saddle point itself can be done with the leading
terms in $X$ since an error of order $\epsilon$ in the location of the
saddle point induces an error of order $\epsilon^2$ in the entropy.
At this level, we must also include the contribution to the entropy
coming from the $\wt\rho$, $\wt\sigma$ integration around the saddle
point:
\be \label{e45.3}
   -{1\over 2} \ln \left|\det \pmatrix{{\p^2 X/ \p \wt\rho^2} &
       {\p^2 X/ \p \wt\rho \p \wt\sigma}\cr
       {\p^2 X/\p \wt\rho \p \wt\sigma} & {\p^2 X
         / \p \wt\sigma^2}}\right|
   \simeq -\ln |Q_e\cdot Q_m| + \ln({1\over 4}+\wt\rho\wt\sigma)
   +\hbox{constant}\, .
\ee
On the other hand, to order $Q^0$, we have
\be \label{e45.4}
   \ln\Delta(\wt\rho,\wt\sigma) \simeq
   \ln |Q_e\cdot Q_m| - \ln({1\over 4}+\wt\rho\wt\sigma)
   +\hbox{constant}\, .
\ee
Thus we see that to this order the contributions \refb{e45.3}
and \refb{e45.4} cancel exactly, leaving behind the contribution
to the entropy
\ben \label{e45.5}
   S_{stat} &\simeq& i\pi X(\wt\rho,\wt\sigma) +  \hbox{constant}
   \nonumber \\
   &=& -i\pi\wt\rho Q_m^2 - i\pi \wt\sigma
   Q_e^2 + 2i\pi
   \Lambda(\wt\rho,\wt\sigma) \, Q_e\cdot Q_m
   \nonumber \\ &&
   +{(k+2) } \ln\wt\sigma -
   \ln f^{(k)}(\rho) -
   \ln f^{(k)}(\sigma) + \hbox{constant} \, ,
\een
evaluated at the saddle point.

Although the saddle point is to be evaluated by extremizing the
leading terms in the expression for $X(\wt\rho,\wt\sigma)$, clearly to
this order we could also use the full expression for $X$ to determine
the location of the extremum. Thus the entropy $S_{stat}$ can be
regarded as the value of $S_{stat}$ given in \refb{e45.5} at the
extremum of this expression. The 
extremization may be done either by
regarding $\wt\rho$, $\wt\sigma$ as independent variables, or by
regarding $\rho$, $\sigma$ defined via \refb{e45.6} as independent
variables. We shall choose to regard $\rho$, $\sigma$ as independent
variables.  If we now define complex variables $a$ and $S$ through
the equations:
\be \label{e45.7}
   \rho = a + i S, \qquad \sigma= -a + i S\, ,
\ee
then from \refb{e5.1}
\be\label{e5.1c}
\Lambda(\wt\rho,\wt\sigma) = - \frac{1}{2}
\, \frac{\rho -\sigma}{(\rho
  +\sigma)} =\frac{1}{2}
\, i\, {a\over S}\, .
\ee
Notice that \refb{e5.1} has a square root in
the expression for $\Lambda$
and there is an ambiguity in choosing the sign of the
square root. This ambiguity is resolved
with the help of \refb{e3.8a} which tells us
that near $v=0$ we have
\be \label{e5.1a}
   \wt v \simeq
   {\rho \over \rho+\sigma}\, .
\ee
This should be identified with $\wt v_-={1\over 2}
-\Lambda$ since we have assumed that
our contour encloses the pole at $\wt v=\wt v_-$. 
This fixes the sign of $\Lambda$ to be the one given in \refb{e5.1c}.
Eqs.\refb{e45.6}, \refb{e45.5} and \refb{e5.1c}  now give
\ben \label{e45.8}
   S_{stat} &=& {\pi\over 2} \bigg[{a^2 +S^2\over S} Q_m^2 +
   {1\over S} Q_e^2 - 2 \, {a\over S}\,  Q_e\cdot Q_m 
   +128\, \pi \, \phi(a,S) \bigg]\nonumber \\
   && + \hbox{constant} + \OO(Q^{-2})\, ,
\een
where
\be \label{ephias}
   \phi(a,S) = -{1\over 64\pi^2} \, 
   \bigg\{ (k+2)\, \ln S +\ln f^{(k)}(a+iS)
   + \ln f^{(k)}(-a+iS) \bigg\} 
\ee
The values of $a$ and $S$ are to be determined by extremizing
\refb{e45.8} with respect to $a$ and $S$. This gives rise to four
equations for four real parameters coming from $a$ and $S$.  For
generic values of $a$ and $S$, $\phi(a,S)$ is complex valued.   
However, for
real values of $a$ and $S$, 
$\phi(a,S)$ is real due to the identity $f^{(k)}(-a+iS)^*=
f^{(k)}(a+iS)$ for real $a$, $S$, and hence 
the expression for $S_{stat}$ given in \refb{e45.8}  is real.  
Thus by restricting $a$ and $S$ to be real
we  end up with two equations
for two real parameters whose solution gives the  
saddle point values of $a$ and $S$ on the real axis.  
Substituting these
values of $a$ and $S$ in \refb{e45.8} we get the value of the
statistical entropy $S_{stat}$. This matches
exactly the entropy of an extremal dyonic black hole given
in eq.(4.11) of \cite{0508042}. The specific form of $\phi(a,S)$ given
in \refb{ephias} agrees with the explicit construction of the
effective action of CHL models as given in appendix B of
\cite{0502126} (see also \cite{9708062}). Thus we see that in the
large charge limit the degeneracy formula for the dyon, given in
\refb{e7}, correctly produces the entropy of the black hole carrying
the same charges, not only to the leading order but also to the first
non-leading order in the inverse power of the charges.\footnote{Note
  that up to this order the black hole entropy agrees with the entropy
  calculated using microcanonical ensemble, \i.e.\ with the logarithm
  of the degeneracy of states.}

\bigskip

\noindent{\bf Acknowledgement:} We wish   
to thank A.~Dabholkar, J.~David, D.~Ghoshal, R.~Gopakumar, S.~Gun,
B.~Ramakrishnan and D.~Suryaramana for useful discussions.

\appendix

\sectiono{(Weak) Jacobi  forms of $\Gamma_1(N)$} \label{sb}

In this appendix we shall review the construction and some
properties of (weak) Jacobi forms of $\Gamma_1(N)$ following
\cite{eichler,mod,skor}. These will be used in appendix \ref{sa} in the
construction of the modular form $\Phi_k$ of the subgroup $G$ of
$Sp(2,\ZZZ)$.

We begin by defining, following \cite{skor},
\be \label{esa1ear}
   \phi_{k,1}(\tau, z) = \eta(\tau)^{-6}\,
   f^{(k)}(\tau)\,  \vt_1(\tau,z)^2  =(\eta(\tau))^{k-4}\,
   (\eta(N\tau))^{k+2} \, \vt_1(\tau,z)^2  \, ,
\ee
where
\be \label{etheta}
   \vt_1(\tau, z) = \sum_{n\in \zzz} \, (-1)^{n-{1\over2}}
   e^{i\pi\tau (n+{1\over 2})^2} e^{2\pi iz(n+{1\over 2})}\, ,
\ee
is a Jacobi theta function. {}From \refb{esa1ear}, \refb{etheta}
we get
\be \label{esa1}
   \phi_{k,1}(\tau, z) = \eta(\tau)^{-6}\,
   f^{(k)}(\tau)\,  \sum_{r,s\in \zzz\atop
     r-s=odd} (-1)^r e^{i\pi\tau(s^2+r^2)/2} e^{2\pi irz}\, .
\ee
{}From \refb{esa1} it follows that if
\be \label{esk1a}
   f^{(k)}(\tau)\eta(\tau)^{-6} = \sum_{n\ge 1} f^{(k)}_n e^{2\pi i \tau
     (n-{1\over 4})} \, ,
\ee
then\footnote{Notice that we have dropped the requirement 
$r-s$=odd that appeared in  
\refb{esa1} since this follows 
from the relation $4l-r^2=4n+s^2-1$
and $l ,n \in \ZZZ$.}
\be \label{esa2}
   \phi_{k,1}(\tau,z) = \sum_{l,r\in \zzz\atop r^2< 4l, l\ge 1}
   C(4l-r^2) e^{2\pi il\tau}e^{2\pi irz}\, ,
\ee
where
\be \label{esa3}
   C(m) = (-1)^m \sum_{s,n\in\zzz \atop n\ge 1} f^{(k)}_n
   \delta_{4n+{s^2-1 },{m }}\, .
\ee
Note that $C(m)$ vanishes for $m\le 0$.   

{}From the definition \refb{esa1},  the modular transformation
law of $\vt_1$, and the fact that $f^{(k)}(\tau)$
transforms as a cusp form of weight $(k+2)$ under $\Gamma_1(N)$,
it follows that
\ben \label{ejac4}   
 &&  \phi_{k,1}\left({a\tau+b\over c\tau+d}, {z\over c\tau+d}\right)
   = (c\tau+d)^k \, \exp(2\pi i c z^2 / (c\tau+d))
   \phi_{k,1}(\tau,z) \nonumber  \\
&& \qquad   \qquad  \qquad   \qquad\qquad   \qquad
\hbox
   {for}\,  \pmatrix{a & b\cr c & d}\in \Gamma_1(N)\, ,
   \nonumber \\
&&   \phi_{k,1}(\tau, z+\lambda\tau + \mu)
   = \exp\left(-2\pi i (\lambda^2\tau + 2\lambda z)\right)
   \phi_{k,1}(\tau, z)  \quad \hbox{for} \quad
    \lambda,\mu\in \ZZZ\, .
   \nonumber \\
\een   
This shows that $\phi_{k,1}(z,\tau)$ transforms as a weak Jacobi 
form of $\Gamma_1(N)$ of weight $k$ and index 1\cite{eichler}.

It also follows from the definition of a cusp form and that
$f^{(k)}(\tau)$ is a cusp form of weight $(k+2)$ of $\Gamma_1(N)$ that
$(c\tau+d)^{-k-2} f^{(k)}((a\tau+b)/(c\tau+d))$ for any element
$\pmatrix{a & b\cr c & d}$ of $SL(2,Z)$ has a series expansion
involving strictly positive powers of $e^{2\pi i\tau}$.  These powers
are integral multiples of $1/N$.  We now introduce the coefficients
$\wh f^{(k)}_n$ through the expansion:
\be \label{ewhf}
\eta(\tau)^{-6} \, (c\tau+d)^{-k-2}\, 
f^{(k)}((a\tau+b)/(c\tau+d)) =\sum_{ n>0\atop nN\in \zzz }\, 
\wh f^{(k)}_n(a,b,c,d)
e^{2\pi i \tau(n-{1\over 4})}\, .
\ee
{}From this and the modular transformation properties and series 
expansion of $\eta(\tau)$ and $\vt_1(\tau,z)$ it then follows, in 
a manner similar to the one that led to eq.\refb{esa2}, that
\ben \label{ecs2}
\wt \phi_{k,1}(\tau,z)
&\equiv&(c\tau+d)^{-k} \exp(-2\pi i c z^2 / (c\tau+d))
\phi_{k,1}((a\tau+b)/(c\tau+d), z/(c\tau+d)) \nonumber \\
&=& \sum_{Nl,r\in \zzz\atop r^2\le 4l+1 - {4\over N}, l>0}
   \wh C(4l-r^2;a,b,c,d) e^{2\pi i l\tau+ 2\pi i r z}\, ,
\qquad \hbox{for} \quad
\pmatrix{a & b\cr c & d}\in SL(2,\ZZZ)\, , \nonumber \\
\een
where
\be \label{edefcht}
\wh C(m;a,b,c,d) = \sum_{s, nN\in \zzz\atop n>0}
(-1)^{s+1} \, \wh f^{(k)}_n(a,b,c,d) \, \delta_{4n+s^2-1,m}\, .
\ee
{}From \refb{edefcht} it follows that $\wh C(m;a,b,c,d)$
vanishes for $m<{4\over N}-1$. For $N=1,2,3$ this restricts $m$
to strictly positive values. Eqs.\refb{ejac4} and \refb{ecs2} 
then shows that $\phi_{k,1}(\tau,z)$ is actually a Jacobi cusp 
form\cite{eichler}. For $N=5,7$ the argument $m$ of $\wh C$ can be
negative, and thus $\phi_{k,1}$ is only a weak Jacobi form.

For $N=1,2,3$ and 5, the values of $k$ are even and $f^{(k)}(\tau)$ is
actually a cusp form of the bigger group $\Gamma_0(N)$\cite{mod}
defined in \refb{esd2}.  For $N=7$ the value of $k$ is odd and hence
under a $\Gamma_0(N)$ transformation $f^{(k)}(\tau)$ transforms as a
cusp form only up to a sign. This can be seen by considering the
$\Gamma_0(7)$ matrix $\pmatrix{-1 & 0 \cr 0 & -1}$ for which the two
sides of the first equation in \refb{fcusp1} differ by $-1$. In
general $f^{(k)}(\tau)$ transforms under $\Gamma_0(N)$ as:
\be \label{egon}
   f^{(k)}\left( {a\tau + b\over
       c\tau +d}\right) = \{\chi(a)\}^{-1}\, (c\tau + d)^{k+2}
   f^{(k)}(\tau)\,  \quad \hbox{for} \quad
   \pmatrix{a & b\cr c & d}\in \Gamma_0(N)\, ,
\ee
where $\chi(a)=1$ or $-1$ determined by the value of $a$ mod
$N$\cite{mod}.  It follows easily from \refb{egon} and that $\chi(a)$
depends only on $a$ mod $N$, that
\be \label{ecapp}
   \chi(a)\, \chi(a') = \chi(aa')\, .
\ee
$\chi(a)$ describes a homomorphism map from $\Gamma_0(N)$ to $\ZZZ_2$
and is known as a Dirichlet character mod 2 of $\Gamma_0(N)$.  For
$N=1,2,3,5$ we have $\chi(a)=1$ for all $a$ since $f^{(k)}(\tau)$ is
actually a cusp form of $\Gamma_0(N)$.  For $N=7$ we know from the
action of $\pm I$ on $f^{(k)}(\tau)$ that $\chi(-1)=-1$.  The only
possible $\chi(a)$ which is consistent with this and \refb{ecapp} is:
\ben \label{echivaluea}
   \chi(a) &=& 1 \quad \hbox{for} \quad
   a=1,2,4 \, \, \hbox{mod} \, 7\, , \nonumber \\
   &=& -1 \quad \hbox{for} \quad
   a=3,5,6 \, \, \hbox{mod} \, 7\, .
\een
{}From \refb{egon} and the modular transformation properties
of $\eta(\tau)$, $\vt_1(\tau,z)$ it follows that
\ben \label{ephigam}
   \phi_{k,1}\left({a\tau+b\over c\tau+d}, {z\over c\tau+d}\right)
   &=& (\chi(a))^{-1}\,
   (c\tau+d)^k \, \exp(2\pi i c z^2 / (c\tau+d)) \phi_{k,1}(\tau,z)
   \nonumber \\ &&
   \qquad \hbox{for} \quad \pmatrix{a & b\cr c & d}\in \Gamma_0(N)
   \, .
\een

We shall now construct a family of other (weak) Jacobi forms of weight
$k$ and index $m$ by applying appropriate operators (known as Hecke
operators) on $\phi_{k,1}(\tau,z)$\cite{eichler,mod}.  Let $S_m$ be
the set of $2\times 2$ matrices satisfying the following relations:
\be \label{ephi1}
   \pmatrix{\alpha & \beta\cr \gamma & \delta}
   \in S_m \quad \hbox{if} \quad
   \alpha, \beta, \gamma, \delta\in \ZZZ, \quad
   \gamma = \hbox{0 mod $N$}, \quad
   \hbox{g.c.d.}(\alpha,N) = 1\, , \quad \alpha\delta-
   \beta\gamma=m\, .
\ee
The set $S_m$ has the property that it is invariant under left
multiplication by any element of $\Gamma_0(N)$.  For proving this we
need to use the fact that for a matrix $\pmatrix{a & b\cr c & d}\in
\Gamma_0(N)$, $a$ satisfies the condition $\hbox{g.c.d.}(a,N)=1$.
This follows automatically from the conditions $ad-bc=1$ and
$c=\hbox{0 mod $N$}$.

We denote by $\Gamma_0(N)\backslash S_m$ the left
coset of $S_m$ by $\Gamma_0(N)$ and define:
\be \label{ephi2}
   \phi_{k,m}(\tau, z) = m^{k-1}\, \sum_{{\pmatrix{\small
         \alpha & \beta\cr \gamma & \delta}}
     \in \Gamma_0(N)\backslash S_m} \, \chi(\alpha)
   (\gamma\tau+\delta)^{-k} \, e^{- 2\pi i m \gamma z^2
     / (\gamma\tau + \delta)} \phi_{k,1}\left(
     {\alpha\tau + \beta \over \gamma\tau + \delta}, {m z \over
       \gamma\tau + \delta}\right)\, .
\ee
{}For this definition of $\phi_{k,m}$ to be sensible, it must be
independent of the representative matrix $\pmatrix{\alpha & \beta\cr
  \gamma & \delta}$ that we choose to describe an element of
$\Gamma_0(N)\backslash S_m$. A different choice $\pmatrix{\wh\alpha &
  \wh\beta\cr \wh\gamma & \wh\delta}$ of the representative element,
corresponding to multiplying $\pmatrix{\alpha & \beta\cr \gamma &
  \delta}$ from the left by an element $\pmatrix{a & b\cr c &
  d}\in\Gamma_0(N)$, amounts to replacing the summand in \refb{ephi2}
by
\ben \label{ereplace1}
&&\chi(\wh\alpha) (\wh\gamma\tau+\wh\delta)^{-k} \, 
e^{- 2\pi i m \wh\gamma z^2
     / (\wh\gamma\tau + \wh\delta)} \phi_{k,1}\left(
     {\wh\alpha\tau + \wh\beta \over \wh\gamma\tau + 
     \wh\delta}, {m z \over
       \wh\gamma\tau + \wh\delta}\right) \nonumber \\
 &=& \chi(\wh\alpha) (c\tau'+d)^{-k}  (\gamma\tau+\delta)^{-k} \,
 e^{-2\pi i c
z^{\prime2} /(c\tau'+d)}
 e^{- 2\pi i m \gamma z^2
     / (\gamma\tau + \delta)} \phi_{k,1}\left(
     {a\tau'+b\over c\tau'+d},{z'\over c\tau'+d}\right) \nonumber \\
     && \qquad \qquad 
     \tau'\equiv{\alpha\tau+\beta\over \gamma\tau+\delta}, \quad
     z'\equiv{mz\over \gamma\tau+\delta}\, .
\een
Using eq.\refb{ephigam} and the relation $\chi(\wh\alpha)
=\chi(\alpha)\chi(a)$ that follows from \refb{ecapp}, we can easily
show that \refb{ereplace1} is equal to the summand in \refb{ephi2}.
Hence the right hand side of \refb{ephi2} is indeed independent of the
representative element of $\Gamma_0(N)\backslash S_m$ that we use for
the computation.

It is also straightforward to study the transformation laws of
$\phi_{k,m}$ under a modular transformation by an element of
$\Gamma_1(N)$, and under shift of $z$ by $(\lambda\tau+\mu)$ for
integer $\lambda$, $\mu$. First consider the effect of modular
transformation of the form
\be \label{enew1}
   \tau \to {a\tau+b\over c\tau+d}, \qquad z\to {z\over c\tau+d},
   \qquad \pmatrix{a & b\cr c & d}\in \Gamma_1(N)\, ,
\ee
for which $\chi(a)=1$.
Defining
\be \label{enew2}
   \pmatrix{\wt\alpha & \wt\beta\cr \wt \gamma & \wt\delta}
   = \pmatrix{\alpha & \beta\cr \gamma & \delta}
   \pmatrix{a & b\cr c & d}\, ,
\ee
for which $\chi(\wt\alpha)=\chi(\alpha)\chi(a)=\chi(\alpha)$,
one can show using \refb{ephi2}, \refb{ejac4} that
\ben \label{enew3}
   && \phi_{k,m}\left({a\tau+b\over c\tau+d}, {z\over c\tau+d}\right)
   = (c\tau+d)^k \, \exp(2\pi i m
   c z^2 / (c\tau+d))  \nonumber \\
   &&  \quad \times m^{k-1}
   \sum_{{\pmatrix{\small
         \alpha & \beta\cr \gamma & \delta}}
     \in \Gamma_0(N)\backslash S_m}\, \chi(\wt\alpha)\,
   (\wt\gamma\tau+\wt\delta)^{-k} \, e^{- 2\pi i m
     \wt\gamma z^2
     / (\wt\gamma\tau + \wt\delta)} \phi_{k,1}\left(
     {\wt\alpha\tau + \wt\beta \over
       \wt\gamma\tau + \wt\delta}, {m z \over
       \wt\gamma\tau + \wt\delta}\right)\, . \nonumber \\
\een
Since the sum over $\pmatrix{\alpha & \beta\cr \gamma & \delta}$ runs
over all the representative elements of $\Gamma_0(N) \backslash S_m$,
we can reinterpret the sum over $\pmatrix{\alpha & \beta \cr \gamma &
  \delta}$ as a sum over inequivalent choices of $\pmatrix{\wt\alpha &
  \wt\beta\cr \wt \gamma & \wt\delta}$. Comparing \refb{enew3} with
\refb{ephi2} we then get
\be \label{enew4}
   \phi_{k,m}\left({a\tau+b\over c\tau+d}, {z\over c\tau+d}\right)
   = (c\tau+d)^k \, \exp(2\pi i m
   c z^2 / (c\tau+d)) \phi_{k,m}(\tau, z)\, .
\ee
Also, using \refb{ejac4} and \refb{ephi2} one can easily show that
\be \label{enew5}
   \phi_{k,m}(\tau, z+\lambda\tau + \mu)
   = \exp\left(-2\pi i m(\lambda^2\tau + 2\lambda z)\right)
   \phi_{k,m}(\tau, z)\, , \quad \lambda,\mu\in \ZZZ\, .
\ee   
This shows that $\phi_{k,m}(\tau, z)$ transforms as a weak Jacobi
form of weight $k$ and index $m$ under $\Gamma_1(N)$.

As an aside, we note that under a $\Gamma_0(N)$ transformation
$\phi_{k,m}$ transforms as
\be \label{ess1}
    \phi_{k,m}\left({a\tau+b\over c\tau+d}, {z\over c\tau+d}\right)
   = (\chi(a))^{-1}(c\tau+d)^k \, \exp(2\pi i m
   c z^2 / (c\tau+d)) \phi_{k,m}(\tau, z)\, .
\ee
This can be proven easily by keeping track of the $\chi(a)$
factors in \refb{enew3}, \refb{enew4} instead of setting it to 1.

We shall now show that we can label the representative
elements of $\Gamma_0(N)\backslash S_m$ by matrices of
the form:
\be \label{ephi3}
   \pmatrix{\alpha & \beta\cr 0 & \delta}, \quad
   \alpha,\beta,\delta\in \ZZZ, \quad \alpha\delta=m, \quad \alpha>0,
   \quad
   \hbox{g.c.d.}(\alpha,N)=1, \quad
   0\le \beta \le \delta - 1\, .
\ee
For this let us consider the product:
\be\label{ephi3a}
   \pmatrix{\alpha' & \beta'\cr
     \gamma' & \delta'}\equiv \pmatrix{a & b\cr c & d} \,
   \pmatrix{\alpha & \beta\cr
     \gamma & \delta}, \qquad \pmatrix{a & b\cr c & d}\in
   \Gamma_0(N), \quad \pmatrix{\alpha & \beta\cr
     \gamma & \delta} \in S_m\, .
\ee
This gives:
\be \label{ephi3b}
   \gamma' = c \alpha + d \gamma\, .
\ee
If we now choose
\be \label{ephi3d}
   c  = -{\gamma \over \hbox{g.c.d.}(\alpha,\gamma)}, \qquad
   d = {\alpha \over \hbox{g.c.d.}(\alpha,\gamma)}\, ,
\ee
we get
\be \label{ephi3e}
   \gamma'=0\, .
\ee
Since $\gamma$ is a multiple of $N$, and since $\alpha$ and $N$ do not
have any common factor, it follows that $c$ given in \refb{ephi3d} is
a multiple of $N$. Furthermore from \refb{ephi3d} it follows that $c$
and $d$ do not have any common factor. Hence it is always possible to
find integers $a$ and $b$ satisfying $ad-bc=1$.  This shows that by
multiplying $\pmatrix{\alpha & \beta\cr \gamma & \delta}$ by an
appropriate $\Gamma_0(N)$ matrix $\pmatrix{a & b\cr c & d}$ from the
left we can bring it to the form
\be \label{ephi3f}
   \pmatrix{\alpha' & \beta'\cr
     0 & \delta'}\, .
\ee
If $\alpha'$ is negative, we can multiply \refb{ephi3f} from the left
by $\pmatrix{-1 & 0\cr 0 & -1}$ to make $\alpha'$ positive. 
We can also multiply \refb{ephi3f} from the left 
by the $\Gamma_0(N)$
matrix
\be \label{ephi3g}
   \pmatrix{1 & k\cr 0 & 1}\, , \qquad k\in \ZZZ\, ,
\ee
to  transform
\be \label{ephi3h}
   \beta' \to \beta'+k\, \delta'\, ,
\ee
preserving the $\gamma'=0$ and $\alpha'>0$ conditions.
Thus by choosing $k$ appropriately we can bring $\beta'$ in the range
$0\le \beta'\le \delta'-1$. Since the final matrix $\pmatrix{\alpha' &
  \beta'\cr 0 & \delta'}$ must still be an element of $S_m$, it must
have determinant $m$ and g.c.d.($\alpha',N)=1$.  This establishes that
the representative elements of $\Gamma_0(N)\backslash S_m$ can be
chosen as in \refb{ephi3}.

Using \refb{ephi3} and \refb{esa2} we can
 rewrite \refb{ephi2} as
\ben \label{eqno}
   \phi_{k,m}(\tau, z) &=& m^{k-1}\,
   \sum_{\alpha,\delta\in\zzz;\alpha>0\atop
     \alpha\delta=m, \, {\small \rm g.c.d.}(\alpha,N)=1}
   \, \chi(\alpha) \, \delta^{-k}\,
   \sum_{\beta=0}^{\delta-1}
   \, \phi_{k,1}((\alpha\tau+\beta)\delta^{-1},
   mz\delta^{-1})\cr
   &=&   m^{k-1}\, \sum_{\alpha,\delta\in\zzz;\alpha>0\atop
     \alpha\delta=m, \, {\small \rm g.c.d.}(\alpha,N)=1}
   \, \chi(\alpha) \, \delta^{-k}\, \sum_{\beta=0}^{\delta-1} \,
   \sum_{n,r\in \zzz\atop n\ge 1, r^2< 4n} C(4n-r^2)
   e^{2\pi i ( n\delta^{-1}(\alpha\tau+\beta) +
     r\delta^{-1} m z)}\, . \nonumber \\
\een
For fixed $\alpha$, $n$, $r$,
the sum over $\beta$  is equal to $\delta$ if
$n= 0$ mod $\delta$, and vanishes otherwise.
Thus we get
\be \label{ephi4}
   \phi_{k,m}(\tau, z)  = m^{k-1}\,
   \sum_{\alpha,\delta\in\zzz;\alpha>0\atop
     \alpha\delta=m, \, {\small \rm g.c.d.}(\alpha,N)=1}
   \, \chi(\alpha)
   \, \delta^{1-k} \, \sum_{n,r\in \zzz\atop
     n\ge 1, r^2< 4n, n\delta^{-1}\in \zzz} C(4n-r^2)
   e^{2\pi i ( n\delta^{-1}\alpha\tau  +
     r\delta^{-1} m z)} \, . \nonumber \\
\ee
We now define:
\be \label{ephi5}
   n'=n\alpha/ \delta, \qquad r'=m r / \delta\, .
\ee
Since $n\delta^{-1}\in \ZZZ$, $m=\alpha\delta$ and
$n,m\ge 1$
we see that
\be \label{ephi6}
   n', r'\in \ZZZ, \qquad \alpha |(m,n',r'), \qquad n'\ge 1\, .
\ee
Furthermore \refb{ephi6} is sufficient to find integers $n\ge 1$, $r$
satisfying \refb{ephi5}. Thus we can replace the sum over $n$ and $r$
in \refb{ephi4} by sum over $n'$ and $r'$ subject to the restriction
given in \refb{ephi6}:
\be \label{ephi7}
   \phi_{k,m}(\tau, z)= \sum_{n',r'\in \zzz\atop n'\ge 1,
     r^{\prime2}< 4 m n'}
   \, \exp(2\pi i(n'\tau+r' z)) \, \sum_{\alpha\in \zzz;\alpha>0\atop
     \alpha|(m,n',r'), \, {\small \rm g.c.d.}(\alpha,N)=1}
   \, \chi(\alpha)
   \, \alpha^{k-1} \, C\left( {4mn'-r^{\prime 2}\over
       \alpha^2} \right)\, .
\ee
This can be rewritten as
\be \label{ejac2}
   \phi_{k,m}(\tau, z) =     \sum_{n,r\in\zzz\atop
     n\ge 1, \, r^2< 4mn}\, a(n,m,r)
   \, e^{2\pi i (n\tau+r z)}   \, ,
\ee
where
\be \label{esk4a}
   a(n,m,r) = \sum_{\alpha\in \zzz;\alpha>0\atop\alpha|(n,m,r), \,
     {\small \rm g.c.d.}(\alpha,N)=1} \chi(\alpha)\,
   \alpha^{k-1}\, C\left({4mn
       -r^2\over \alpha^2}\right)\, .
\ee   
{}Since $C(s)$ vanishes for $s\le 0$, we have
\be \label{elight}
a(n,m,r) = 0\, , \qquad \hbox{for} \quad r^2\ge 4mn\, .
\ee

\sectiono{Proof of modular property of $\Phi_k$}
\label{sa}

In this appendix we shall prove, following \cite{eichler,skor},
that $\Phi_k$ defined in
\refb{eisk3}:
\ben \label{esk3a}
   \Phi_k(\rho,\sigma, v) = \sum_{n,m,r\in\zzz\atop
     n,m\ge 1, \, r^2< 4mn}\, a(n,m,r)
   \, e^{2\pi i (n\rho+m\sigma+rv)}
   \, ,
\een
transforms as a modular form of weight $k$ under the group $G$ defined
in \refb{egroup}, and also that for small $v$ it has the factorization
property described in \refb{e6}.\footnote{Note that due to
  eq.\refb{elight}, the exponent appearing in \refb{esk3a} has
  strictly negative real part for $(\rho,\sigma,v)$ satisfying
  \refb{elight1}, and hence the Fourier expansion \refb{esk3a} is
  sensible in the region \refb{elight1} for large imaginary values of
  $\rho$ and $\sigma$.}  We use \refb{ejac2} to rewrite \refb{esk3a}
as
\be \label{ejac3}
   \Phi_k(\rho,\sigma,v) =\sum_{m\ge 1} \phi_{k,m}(\rho, v)
   e^{2\pi i m\sigma}\, ,
\ee and study the transformation law of $\Phi_k(\rho,\sigma,v)$
under various special $Sp(2,\ZZZ)$ transformations:
\be
\label{espr1}
   \Omega\to (A\Omega+B)(C\Omega+D)^{-1}, \qquad
   \Omega\equiv \pmatrix{\rho & v \cr v & \sigma}\, .
\ee
\begin{enumerate}
\item First consider the transformation generated by
\be \label{ejac6}
   \pmatrix{A & B \cr C & D} = \pmatrix{a & 0 & b & 0\cr
     0 & 1 & 0 & 0\cr c & 0 & d & 0\cr 0 & 0 & 0 & 1}\, , \quad
   ad-bc=1, \quad
   \quad c = \hbox{0 mod $N$}\, , \quad a,d=
   \hbox{1 mod $N$}\, .
\ee
This corresponds to
\ben \label{ejac7}
&&   \rho\to \rho'={a\rho+b\over c\rho+d}, \qquad
   v \to v'={v\over c\rho+d}\, , \qquad \sigma\to \sigma'=
   \sigma - {cv^2\over c\rho + d}\, , \nonumber \\
&&   \det(C\Omega+D) = (c\rho+d)\, .
\een
Hence it follows from \refb{ejac3},  
\refb{enew4}
that
\be \label{ejac8}
   \Phi_k(\rho',\sigma',v')= \det(C\Omega+D)^k
   \Phi_k(\rho,\sigma,v)\, .
\ee
As an aside we note that if $\pmatrix{a & b \cr c & d}\in
\Gamma_0(N)$, then due to \refb{ejac3},  
\refb{ess1},
\be \label{egamp}
\Phi_k(\rho',\sigma', v')= (\chi(a))^{-1} \det(C\Omega+D)^k
   \Phi_k(\rho,\sigma,v)\, .
\ee

\item Next consider the transformation:
\be \label{ejac9}
   \pmatrix{A & B \cr C & D} = \pmatrix{1 & 0 & 0 & \mu\cr
     \lambda & 1 & \mu & 0\cr 0 & 0 & 1 & -\lambda\cr
     0 & 0 & 0 & 1}\, , \qquad \lambda, \mu \in \ZZZ\,.
\ee
which induces the transformation:
\be \label{ejac7a}
   \rho\to \rho'=\rho, \qquad
   v \to v'=v+\lambda\rho+\mu\, , \qquad \sigma\to \sigma'=
   \sigma+2\lambda v+\lambda^2 \rho +\lambda\mu\, .
\ee
It follows from \refb{ejac3},   \refb{enew5}
that under this transformation:
\be \label{ejac8a}
   \Phi_k(\rho',\sigma',v')= \Phi_k(\rho,\sigma, v)
   = \det(C\Omega+D)^k
   \Phi_k(\rho,\sigma,v)\, ,
\ee
since $\det(C\Omega+D)=1$.
\item The matrix
\be \label{ejac10}
   \pmatrix{A & B \cr C & D} = \pmatrix{0 & 1 & 0 & 0\cr
     -1 & 0 & 0 & 0\cr 0 & 0 & 0 & 1\cr 0 & 0 & -1 & 0}\, ,
\ee
induces the transformation
\be \label{ejac11}
   \rho\to \rho'=\sigma, \qquad \sigma\to \sigma'=\rho, \qquad
   v\to v'=-v\, .
\ee
{}From \refb{esk4a}, \refb{esk3a} it follows that
$\Phi_k(\rho,\sigma,v)$ is symmetric under the exchange of
$\rho$ and $\sigma$ and also under $v\to -v$.
As a result we get, under \refb{ejac11}
\be \label{ejac8c}
   \Phi_k(\rho',\sigma',v')= \Phi_k(\rho, \sigma,v)
   = \det(C\Omega+D)^k
   \Phi_k(\rho,\sigma,v)\, ,
\ee
since $\det(C\Omega+D)=1$.

\end{enumerate}

This shows that $\Phi_k(\rho,\sigma,v)$ transforms as a modular form
of weight $k$ under the transformations \refb{ejac6}, \refb{ejac9} and
\refb{ejac10}. Since these transformations generate the group $G$, it
follows that $\Phi_k$ transforms as a modular form of weight $k$ under
the entire group $G$.

{}From \refb{egamp} and the $\rho\leftrightarrow\sigma$ exchange
symmetry \refb{ejac11} it also follows that under an $Sp(2,\ZZZ)$
transformation:
\be \label{eyy1}
\pmatrix{A & B\cr C & D}= \pmatrix{
a & 0 & b & 0\cr 0 & a & 0 & -b\cr c & 0 & d & 0\cr 0 & -c & 0 &
d}\, , \qquad \pmatrix{ a & b\cr c & d}\in \Gamma_0(N)\, ,
\ee
$\Phi_k$ transforms as \be \label{eyy2} \Phi_k\left( (A\Omega+ B)
(C\Omega+ D)^{-1}\right) = \det(C\Omega + D)^k \Phi_k(\Omega)\, ,
\ee since the $\chi(a)$ factors arising from the $\Gamma_0(N)$
transformations acting on $\rho$ and $\sigma$ cancel.

We shall now turn to the study of $\Phi_k$ for small $v$ and
indicate the proof of eq.\refb{e6}. For this we note from
\refb{esa1ear} and the relation
\be \label{ere1}
   \vt_1(\tau, z) \simeq 2\pi \eta(\tau)^3 z + \OO(z^3)\, ,
\ee
that
\be \label{ere2}
   \phi_{k,1}(\tau, z) = 4\pi^2 \, f^{(k)}(\tau) \, z^2 + \OO(z^4)\, ,
\ee
for small $z$. Using eq.\refb{ephi2}  we now see that each
$\phi_{k,m}(\tau,z)$  also vanishes as $z^2$ for small $z$.
\refb{ejac3} then gives, for small $v$,
\be \label{ere3}
   \Phi_k(\rho,\sigma,v) = v^2 \, F_k(\rho, \sigma) + \OO(v^4)\, ,
\ee
where
\be \label{efdef}
F_k(\rho,\sigma) = 4\pi^2 \, \sum_{m\ge 1} e^{2\pi im\sigma}
\, m^{k+1} \, \sum_{\pmatrix{\alpha & \beta\cr \gamma & \delta}
\in \Gamma_0(N)\backslash S_m} \chi(\alpha) (\gamma\rho+\delta)^{-k-2}
\, f^{(k)}\left({\alpha\rho+\beta\over \gamma\rho+\delta}  \right)\, .
\ee
It  follows from the modular transformation property of
$\Phi_k(\rho,\sigma,v)$ that
\ben \label{ere4}
   && F_k((a\rho+b)(c\rho+d)^{-1}, \sigma) = (c\rho+d)^{k+2}
   F_k(\rho,\sigma)\, , \nonumber \\
   &&
   F_k(\rho, (a\sigma+b) (c\sigma+d)^{-1}) =
   (c\sigma+d)^{k+2}
   F_k(\rho,\sigma),  \nonumber \\
   && \qquad  \qquad \hbox{for} \quad \pmatrix{a & b
     \cr c & d} \in \Gamma_1(N)\, .
\een
Finally, since $f^{(k)}(\tau)$ vanishes at  the cusps in the $\tau$-plane,
it follows from \refb{efdef} that for fixed $\sigma$,
$F_k(\rho,\sigma)$ vanishes at the cusps in the $\rho$ plane.
Due to the $\rho\leftrightarrow\sigma$ symmetry the same result is
true in the $\sigma$-plane for fixed $\rho$. Hence
$F_k(\rho,\sigma)$ vanishes at the cusps in the $\rho$ as well as in
the $\sigma$-plane.

The above results show that for fixed $\sigma$ we can regard
$F_k(\rho,\sigma)$ as a cusp form of $\Gamma_1(N)$ of weight
$k+2$ in the $\rho$-plane. Similarly for fixed $\rho$ we can regard
$F_k(\rho,\sigma)$ as a cusp form of $\Gamma_1(N)$ of weight
$k+2$ in the $\sigma$ plane. Since these cusp forms are known to
be unique and proportional to $f^{(k)}(\rho)$ and $f^{(k)}(\sigma)$
respectively\cite{mod}, we get
\be \label{ere5}
   F_k(\rho,\sigma)=C_0\, f^{(k)}(\rho) f^{(k)}(\sigma)
\ee
where $C_0$ is a constant of proportionality.
By examining the behaviour
of both sides for large imaginary values of $\rho$ and $\sigma$
one can verify that $C_0=4\pi^2$.
This proves \refb{e6}:
\be \label{eproof}
\Phi_k(\rho,\sigma,v)\simeq 4\pi^2 \, v^2 \, f^{(k)}(\rho)\, 
f^{(k)}(\sigma)+\OO(v^4)\, .
\ee

\sectiono{Fourier expansion of $\wt\Phi_k$} \label{sc}

In this appendix we shall show that $\wt\Phi_k(\wt\rho, \wt\sigma, \wt
v)$ given in \refb{ewtphi1} has an expansion of the form described in
\refb{e33.1} with integer coefficients $b(q,r,s)$.  For this we use
the Fourier expansion of $\Phi_k$ given in \refb{ejac3} with $\rho$,
$\sigma$ exchanged to express \refb{ewtphi1} as
\be \label{eac1}
   \wt\Phi_k(\wt\rho, \wt\sigma, \wt v) = \sum_{m\ge 1}
   e^{2\pi i m\wt\rho}\, \wt \sigma^{-k} \, e^{-2\pi i m\wt v^2
     /\wt\sigma} \, \phi_{k,m}(-1/\wt\sigma, \wt v /\wt\sigma)\, .
\ee
Thus in order to find the Fourier expansion of
$\wt\Phi_k$ in $\wt\rho$,
$\wt\sigma$ and $\wt v$, we need to find the Fourier expansion of
$\wt\sigma^{-k}\, e^{-2\pi i m\wt v^2
 /\wt\sigma} \,
\phi_{k,m}(-1/\wt\sigma, \wt v/\wt\sigma)$ in $\wt\sigma$ and
$\wt v$.

We first analyze $\phi_{k,1}$.  {}From eqs.\refb{esa1ear},
\refb{emod}, and the known modular transformation properties of
$\eta(\tau)$ and $\vt_1(\tau,z)$:
\be \label{eetamod}
   \eta(-{1\over \tau}) = (-i\tau)^{1/2} \eta(\tau) \, , \qquad
   \vt_1(-{1\over \tau}, {z\over \tau}) = -i\, (-i\tau)^{1/2}
   e^{i\pi z^2/\tau} \, \vt_1(\tau, z)\, ,
\ee
we see that
\be \label{eac2}
   \phi_{k,1}(-1/\tau, z/\tau) = (i\sqrt N)^{-k-2}\,
   \tau^k \, e^{2\pi i z^2/\tau}\, \eta(\tau)^{-6}\,
   f^{(k)}(\tau/N)\,  \vt_1(\tau,z)^2  \, .
\ee
Thus
\be \label{eac3}
   \wt \sigma^{-k} \, e^{-2\pi i  \wt v^2
     /\wt\sigma} \, \phi_{k,1}(-1/\wt\sigma, \wt v /\wt\sigma)
   = (i\sqrt N)^{-k-2}\, \eta(\wt\sigma)^{-6}\,
   f^{(k)}(\wt\sigma/N)\,  \vt_1(\wt\sigma,\wt v)^2  \, .
\ee
If we define the coefficients $\wt f^{(k)}_n$ through
\be \label{ewtfn}
\eta(\tau)^{-6} f^{(k)}(\tau/N) = \sum_{n>0\atop nN\in \zzz} \wt f^{(k)}_n
e^{2\pi i \tau (n-{1\over 4})}\, ,
\ee
then due to eq.\refb{etheta} 
$\eta(\tau)^{-6}f^{(k)}(\tau/N)\vt_1(\tau, z)^2$ has a
Fourier expansion of the form:
\ben \label{eac3aa}
&&   \eta(\tau)^{-6}\,
   f^{(k)}(\tau/N)\,  \vt_1(\tau,z)^2 =
   \sum_{r,s\in\zzz\atop r\ge 1} \, d(r,s)
   e^{2\pi i r\tau/N + 2\pi i s
     z} \nonumber \\
&\equiv &  -e^{2\pi i\tau/N}
   e^{2\pi i z} \bigg( 1  - 2e^{-2\pi iz} + e^{-4\pi i z}
   + \sum_{r,s\in\zzz\atop r\ge 1}
   c(r,s) e^{2\pi i r\tau/N + 2\pi i s
     z}\bigg) \nonumber \\
 \een
where
\be \label{edrs}
d(r,s) =\wt C(4r/N-s^2)\, ,
\ee
\be \label{ewtcm}
\wt C(m)=\sum_{l,nN\in \zzz\atop n>0}(-1)^{l+1}\, \wt f^{(k)}_n 
\delta_{4n+l^2-1,m}
\, .
\ee
Since $\wt f^{(k)}_n$ are integers,  $c(r,s)$ and $d(r,s)$  
are also integers. Using
\refb{eac3} and \refb{eac3aa} we get
\be \label{eac4}
   \wt \sigma^{-k} \, e^{-2\pi i  \wt v^2
     /\wt\sigma} \, \phi_{k,1}(-1/\wt\sigma, \wt v /\wt\sigma)
   =
    (i\sqrt N)^{-k-2} \sum_{r,s\in\zzz\atop r\ge 1}
   d(r,s) e^{2\pi i r\wt\sigma/N + 2\pi i s
     \wt v}  \, .\nonumber \\
\ee

We now turn to the analysis of $\phi_{k,m}$.
We begin with
\refb{ephi2}:
\be \label{ephi2a}
   \phi_{k,m}(\tau, z) = m^{k-1}\, \sum_{{\pmatrix{\small
         \alpha & \beta\cr \gamma & \delta}}
     \in \Gamma_0(N)\backslash S_m}\, \chi(\alpha) \,
   (\gamma\tau+\delta)^{-k} \, e^{- 2\pi i m \gamma z^2
     / (\gamma\tau + \delta)} \phi_{k,1}\left(
     {\alpha\tau + \beta \over \gamma\tau + \delta}, {m z \over
       \gamma\tau + \delta}\right)\, .
\ee
Here  $\alpha,\beta,\gamma,\delta$ satisfy
\be \label{eacc0}
   \alpha,\beta,\gamma,\delta\in \ZZZ, \qquad
   \alpha\delta - \beta\gamma
   =m, \qquad \gamma=\hbox{0 mod $N$}, \qquad g.c.d.(\alpha,N)
   = 1\, .
\ee
We shall analyze \refb{ephi2a} by choosing appropriate representative
matrices $\pmatrix{\alpha & \beta\cr \gamma & \delta}$. However the
representatives chosen in appendix \ref{sb}, which were useful for
studying the Fourier expansion of $\Phi_k$, will not be useful for
studying the Fourier expansion of $\wt\Phi_k$, and we need to proceed
differently.  First consider the case $m\ne 0$ mod $N$. In this case
we can find a matrix $\pmatrix{a & b\cr c & d}\in \Gamma_0(N)$ such
that
\be \label{eacc1}
   \pmatrix{a & b\cr c & d} \, \pmatrix{
     \alpha & \beta\cr \gamma & \delta} = \pmatrix{\alpha' & 0\cr
     \gamma' & \delta'}\, , \quad \alpha'>0\, .
\ee   
This is done by choosing
\be \label{eacc2}
   a = \mp {\delta \over g.c.d.(\beta, \delta)}, \qquad b = \pm
   {\beta \over g.c.d.(\beta, \delta)}\, ,
\ee
where the overall sign of $a$ and $b$ is chosen so that $
\alpha'=a\alpha
+b\gamma>0$.
As long as $m\ne 0$ mod $N$, the conditions \refb{eacc0}
ensures that $\delta\ne 0$ mod $N$. As a result $a$ determined
from \refb{eacc2} satisfies $a\ne 0$ mod $N$. In this case it is
always possible to choose integers $c$ and $d$ satisfying
\be \label{eacc3}
   ad-bc=1, \qquad c=\hbox{0 mod $N$}\, .
\ee
We shall use this freedom to choose the representative matrices
in the sum in \refb{ephi2a} to be of the form
\be \label{eacc4}
   \pmatrix{\alpha & 0\cr
     \gamma & \delta}, \qquad
   \alpha\delta = m, \qquad
   \alpha>0\, , \qquad \gamma=\hbox{0 mod $N$}, \qquad
   g.c.d.(\alpha,N)=1\, .
\ee
Furthermore using the freedom of multiplying this matrix
from the left by the $\Gamma_0(N)$ matrix
\be \label{eacc4a}
   \pmatrix{ 1 & 0\cr N & 1}\, ,
\ee
which shifts $\gamma$ by $\alpha N$, we can restrict $\gamma$
to be of the form:
\be \label{eacc5}
   \gamma= \gamma_0 N, \qquad 0\le \gamma_0\le \alpha-1\, .
\ee
This gives
\ben \label{eacc6}
   \phi_{k,m}(\tau, z) &=& m^{k-1}\sum_{\alpha, \delta;\alpha>0
   \atop
     \alpha\delta=m, g.c.d.(\alpha,N)=1}
   \chi(\alpha)\,
   \sum_{\gamma_0=0}^{\alpha-1} (\gamma_0 N \tau
   + \delta)^{-k} e^{-2\pi i m \gamma_0 N z^2 / (\gamma_0 N \tau
     +\delta)} \, \nonumber \\
   && \qquad \qquad
   \times \phi_{k,1} \left( {\alpha\tau\over \gamma_0 N\tau+
       \delta}, {mz\over \gamma_0 N\tau+\delta}\right)\, .
\een
The sum over $\alpha$, $\delta$, $\gamma_0$ run over integer values
only. From now on this is the convention we shall be using for all
summation indices unless mentioned otherwise.  We now use \refb{eac2}
to reexpress \refb{eacc6} as
\ben \label{eacc7}
   \phi_{k,m}(\tau, z) &=& (i\sqrt N)^{-k-2}\, m^{k-1}
   \sum_{\alpha, \delta;\alpha>0\atop
     \alpha\delta=m, g.c.d.(\alpha,N)=1}
   \chi(\alpha)\,
   \sum_{\gamma_0=0}^{\alpha-1} (-\alpha\tau)^{-k}
   e^{-2\pi i m z^2/\tau} \nonumber \\
   && \eta^{-6}\left(-{\gamma_0 N\tau+\delta
       \over \alpha\tau}\right) f^{(k)}\left(-{\gamma_0 N\tau+\delta
       \over N\alpha\tau}\right)
   \vt_1^2\left(-{\gamma_0 N\tau+\delta
       \over  \alpha\tau}, -{mz\over \alpha\tau}\right)\, .
\een
Setting $\tau=-1/\wt\sigma$, $z=\wt v/\wt\sigma$ in \refb{eacc7} we get
\ben \label{eacc8}
   && e^{-2\pi i m \wt v^2/\wt\sigma} \, \wt\sigma^{-k} \,
   \phi_{k,m}\left(-{1\over \wt\sigma}, {\wt v\over
       \wt\sigma}\right) \nonumber \\
   &=& (i\sqrt N)^{-k-2} \sum_{\alpha, \delta;\alpha>0\atop
     \alpha\delta=m, g.c.d.(\alpha,N)=1}
   \chi(\alpha)\,
   \sum_{\gamma_0=0}^{\alpha-1} \, \alpha^{-1}
   \delta^{k-1} \eta^{-6}\left( -{\gamma_0 N\over \alpha}
     +{\delta\over \alpha}\, \wt\sigma\right)
   f^{(k)}\left(-{\gamma_0  \over \alpha}
     +{\delta\over N\alpha}\, \wt\sigma\right)
   \nonumber \\
   && \times \, \vt_1^2\left(-{\gamma_0 N\over \alpha}
     +{\delta\over \alpha}\, \wt\sigma, \delta \, \wt v\right) \, .
\een
Using \refb{eac3aa} we can express \refb{eacc8}   as
\ben \label{eacc8a}
   && e^{-2\pi i m \wt v^2/\wt\sigma} \, \wt\sigma^{-k} \,
   \phi_{k,m}\left(-{1\over \wt\sigma}, {\wt v\over
       \wt\sigma}\right) \nonumber \\
   &=& (i\sqrt N)^{-k-2} \sum_{\alpha, \delta;\alpha>0\atop
     \alpha\delta=m, g.c.d.(\alpha,N)=1}
   \chi(\alpha)\,
   \sum_{\gamma_0=0}^{\alpha-1} \, \alpha^{-1}
   \delta^{k-1} \sum_{r,s\in \zzz\atop r\ge 1} \, d(r,s)\,
   e^{2\pi i r(-\alpha^{-1}\gamma_0+\delta\alpha^{-1}
     \wt\sigma/N) + 2\pi i \delta s\wt v}\, . \nonumber \\
\een
Performing the sum over $\gamma_0$ for fixed $\alpha$, $\delta$, $r$
and $s$ we see that the sum vanishes unless $r/\alpha\in \ZZZ$, and is
equal to $\alpha$ if $r/\alpha\in \ZZZ$. Writing $r = r_0\alpha$, we
get from \refb{eacc8a}:\footnote{The condition $r_0\delta>0$ 
in \refb{eacc8b} follows
  from $r_0m/\delta=r_0\alpha=r\ge 1$, and $m\ge 1$.}
\ben \label{eacc8b}
   && e^{-2\pi i m \wt v^2/\wt\sigma} \, \wt\sigma^{-k} \,
   \phi_{k,m}\left(-{1\over \wt\sigma}, {\wt v\over
       \wt\sigma}\right) \nonumber \\
   &=& (i\sqrt N)^{-k-2} \sum_{\alpha, \delta;\alpha>0\atop
     \alpha\delta=m, g.c.d.(\alpha,N)=1}\chi(\alpha)\,
   \delta^{k-1} \sum_{r_0,s\in \zzz\atop r_0\delta>0}\,
   \, d(r_0\alpha,s)\,
   e^{2\pi i\delta r_0
     \wt\sigma/N+ 2\pi i \delta\, s\wt v}\, . \nonumber \\
\een

Finally we turn to the case where $m=0$ mod $N$. In this case
eq.\refb{eacc0} gives $\delta=0$ mod $N$.
We split the sum in \refb{ephi2a} into two parts:
\be \label{ead1}
   \phi_{k,m}(\tau, z) = I_1(\tau,z) + I_2(\tau,z)\, ,
\ee
where $I_1$ and $I_2$ represent the sum over
$(\alpha,\beta,\gamma,\delta)$ restricted as follows:\footnote{It is
  easy to verify that the conditions on $(\beta,\delta)$ appearing in
  \refb{ead2} are preserved under left multiplication of
  $\pmatrix{\alpha & \beta\cr \gamma & \delta}$ by an element of
  $\Gamma_0(N)$ and hence the splitting of the sum into $I_1$ and
  $I_2$ is independent of the choice of representative elements of
  $\Gamma_0(N)\backslash S_m$.}
\ben \label{ead2}
   I_1  &:& {\delta\over g.c.d.(\beta,\delta)}
   \ne \hbox{0 mod $N$}\, , \nonumber \\
   I_2  &:& {\delta\over g.c.d.(\beta,\delta)}
   = \hbox{0 mod $N$}\, .
\een
For $I_1$ we can bring the matrix $\pmatrix{\alpha &\beta\cr
\gamma & \delta}$ into the form given in \refb{eacc4},
\refb{eacc5} using transformations given in
\refb{eacc1}-\refb{eacc3}.
The analysis now proceeds exactly as in the
$m\ne 0$ mod $N$ case, and we get the analog of
\refb{eacc8b}
\ben \label{ead3}
   && e^{-2\pi i m \wt v^2/\wt\sigma} \, \wt\sigma^{-k} \,
   I_1(-1/\wt\sigma,\wt v/\wt\sigma) \nonumber \\
   &=& (i\sqrt N)^{-k-2} \sum_{\alpha, \delta;\alpha>0\atop
     \alpha\delta=m, g.c.d.(\alpha,N)=1} \chi(\alpha) \,
   \delta^{k-1} \sum_{r_0,s\in \zzz\atop r_0\delta>0}\,
   \, d(r_0\alpha,s)\,
   e^{2\pi i\delta r_0
     \wt\sigma/N+ 2\pi i \delta s\wt v}\, . \nonumber \\
\een
For $I_2$ we cannot choose $a$ as in \refb{eacc2} since this will
give $a=0$ mod $N$ and it will be impossible to find integers
$b$, $c$ satisfying \refb{eacc3}. Instead we choose
\be \label{ead4}
   c = - {\delta \over g.c.d.(\beta, \delta)}, \qquad d =
   {\beta \over g.c.d.(\beta, \delta)}\, ,
\ee
to bring the representative matrices in the sum in
\refb{ephi2a} to  the form
\be \label{ead5}
   \pmatrix{\alpha & -\beta \cr \gamma & 0} \, ,
   \qquad \beta\gamma=m, \qquad \beta>0,
   \qquad \gamma=\hbox{0 mod N}
   \qquad g.c.d.(\alpha,N)=1 \, .
\ee
Furthermore using the freedom of multiplying this matrix
from the left by the $\Gamma_0(N)$ matrix
\be \label{ead6}
   \pmatrix{1 & 1\cr 0 & 1}\, ,
\ee
which shifts $\alpha$ by $\gamma$,  we restrict $\alpha$
to the range $0\le\alpha\le \gamma-1$. Defining integers
 $m_0$ and $\gamma_0$ through
\be \label{ead7}
   m_0 = m/N, \qquad \gamma_0=\gamma/ N\, ,
\ee
we get, using eqs.\refb{ephi2a}, \refb{ead2}, \refb{ead5},
\be \label{ead8}
   I_2(\tau,z) = (m_0\, N)^{k-1} \sum_{\beta,\gamma_0;\beta>0\atop
     \beta\gamma_0=m_0}
   \sum_{\alpha=1\atop g.c.d.(\alpha,N)=1}^{\gamma_0N-1}
   \chi(\alpha)\, (\gamma_0 N \tau)^{-k} e^{-2\pi i m z^2/\tau}
   \phi_{k,1} \left({\alpha\tau-\beta\over \gamma_0 N \tau}, {m_0
       z\over
       \gamma_0 \tau}\right)\, .
\ee
We  now set $\tau=-1/\wt\sigma$, $z=\wt v/\wt\sigma$, and
write $\alpha=s+pN$ with $1\le s\le N-1$,
$0\le p\le \gamma_0-1$.
This gives:
\be \label{ead9}
   I_2(-1/\wt\sigma,\wt v/\wt\sigma) = 
   \sum_{\beta,\gamma_0;\beta>0\atop
     \beta\gamma_0=m_0}
   \sum_{s=1}^{N-1}\sum_{p=0}^{\gamma_0-1} \chi(s)
   (-\wt\sigma)^k \beta^{k-1} (\gamma_0N)^{-1} e^{2\pi i m
     \wt v^2
     /\wt\sigma} \, \phi_{k,1}\left( {s+pN\over \gamma_0 N}
     + {\beta \over \gamma_0 N}\wt\sigma, -\beta\wt v  \right)\, .
\ee
Using \refb{esa2} this may be expressed as:
\ben \label{ead10}
   && e^{-2\pi i m \wt v^2/\wt\sigma} \, \wt\sigma^{-k} \,
   I_2(-1/\wt\sigma,\wt v/\wt\sigma) \nonumber \\
   &=& (-1)^k  N^{-1} \sum_{\beta,\gamma_0;\beta>0\atop
     \beta\gamma_0=m_0} \, \beta^{k-1}\, (\gamma_0)^{-1}
   \sum_{s=1}^{N-1}\sum_{p=0}^{\gamma_0-1} \chi(s)
   \sum_{l,r\atop r^2< 4l, l\ge 1}
   C(4l-r^2) e^{2\pi i l \gamma_0^{-1} N^{-1}
     (s+pN+\beta\wt\sigma)
     - 2\pi i r\beta \wt v} \,. \nonumber \\
\een
The sum over $p$ produces a factor
of $\gamma_0$ if $l=l_0\gamma_0 $, $l_0\in \ZZZ$ and
vanishes otherwise.  This gives
\ben \label{ead10aa}
   && e^{-2\pi i m \wt v^2/\wt\sigma} \, \wt\sigma^{-k} \,
   I_2(-1/\wt\sigma,\wt v/\wt\sigma) \nonumber \\
   &=& (-1)^k  N^{-1} \sum_{\beta,\gamma_0;\beta>0\atop
     \beta\gamma_0=m_0} \, \beta^{k-1}\,
   \sum_{l_0,r\atop r^2< 4l_0\gamma_0, l_0\gamma_0\ge 1}
   C(4l_0\gamma_0-r^2) e^{2\pi i l_0    \beta\wt\sigma/N
     - 2\pi i r\beta \wt v}\sum_{s=1}^{N-1} e^{2\pi i l_0 s/N} \chi(s)
   \,. \nonumber \\
\een
In this sum $l_0\beta=l_0m/\gamma=l_0m/\gamma_0N>0$ since
$l_0\gamma_0>0$, $m>0$, $N>0$. Thus $\wt\sigma$ in the exponent always
appears with positive coefficient.  We now consider the cases $N\ne 7$
and $N=7$ separately.  For $N\ne 7$ we have $\chi(s)=1$ and the sum
over $s$ gives $(N-1)$ if $l_0=0$ mod $N$ and $-1$ if $l_0\ne 0$ mod
$N$. Since in this case $k\ge 2$ and even, and $C(s)$ are integers, we
can express $I_2(-1/\wt\sigma,\wt v/\wt\sigma)$ as:
\be \label{ead11}
e^{-2\pi i m \wt v^2/\wt\sigma} \, \wt\sigma^{-k} \,   
I_2(-1/\wt\sigma,\wt v/\wt\sigma)
   =-(i\sqrt N)^{-k-2} \, \sum_{r,p\in \zzz\atop p\ge 1}
   e_N(m_0,p,r) \, e^{2\pi i p\wt\sigma/N + 2\pi i r \wt v}\, ,
\ee
with integer coefficients $e_N(m_0,p,r)$.
For $N=7$, we can use the identity:
\ben \label{eidentity}
   \sum_{s=1}^6 \, e^{2\pi i  l_0 s / 7} \chi(s) &=& 0 \quad
   \hbox{for} \quad l_0=0 \quad \hbox{mod} \, N
   \nonumber \\
   &=&  i\sqrt{7}\,
   \chi(l_0)\quad
   \hbox{for} \quad l_0\ne 0 \quad \hbox{mod} \, N\, ,
\een
which follows from the values of $\chi(a)$ given in \refb{echivaluea}.
Substituting this into \refb{ead10aa}, and noting that $k=1$ for
$N=7$, we see that even in this case we can express
$I_2(-1/\wt\sigma,\wt v/\wt\sigma)$ as:
\be \label{ead11a}
e^{-2\pi i m \wt v^2/\wt\sigma} \, \wt\sigma^{-k} \,
 I_2(-1/\wt\sigma,\wt v/\wt\sigma)=-(i\sqrt N)^{-k-2} \, 
   \sum_{r,p\in \zzz\atop p\ge 1}
   e_7(m_0,p,r) \, e^{2\pi i p\wt\sigma/N + 2\pi i r \wt v}\, ,
\ee
with integer coefficients $e_7(m_0,p,r)$.
 
It is now time to collect all the information together. Using
eqs.\refb{eac1},  \refb{eac4}, \refb{eacc8b}-\refb{ead3},
\refb{ead11}, \refb{ead11a} we get
\ben \label{efirstexp}
&& \wt\Phi_k(\wt\rho,\wt\sigma, \wt v) \nonumber \\
&=&  
(i\sqrt N)^{-k-2} \sum_{m\ge 1}e^{2\pi i m\wt\rho}
\, \sum_{\alpha, \delta;\alpha>0\atop
     \alpha\delta=m, g.c.d.(\alpha,N)=1}\chi(\alpha)\,
   \delta^{k-1} \sum_{r_0,s \atop r_0\delta>0}\,
   \, d(r_0\alpha,s)\,
   e^{2\pi i\delta r_0
     \wt\sigma/N+ 2\pi i \delta\, s\wt v}
\nonumber \\
&& - (i\sqrt N)^{-k-2}\, \sum_{m_0\ge 1} \,
e^{2\pi i m_0 N\wt\rho} \sum_{r,p \atop p\ge 1}
e_N(m_0,p,r) e^{2\pi i p\wt\sigma/N+2\pi i r \wt v}\, .
\een 
In the above expression the sum over the indices $m$, $\alpha$,
$\delta$, $r_0$, $s$, $r$, $p$ run over integer values. Using the
integrality of the coefficients $d(r,s)$, $e_N(m_0,p,r)$ and the
expression for $d(1,s)$ implicit in \refb{eac3aa} we can rewrite this
as
\ben \label{e33.1a}
   \wt\Phi_k(\wt\rho, \wt\sigma, \wt v) &=& C\, e^{2\pi i\wt\rho
     +2\pi i \wt\sigma/N + 2\pi i \wt v} \,
   \bigg( 1 - 2e^{-2\pi i \wt v} + e^{-4\pi i \wt v} \nonumber \\
   &&
   + \sum_{q,r,s\in \zzz\atop q,r\ge 0,q+r\ge 1} \, b(q,r,s)\,
   e^{2\pi i q\wt\rho
     +2\pi i r\wt\sigma/N + 2\pi i s\wt v}\bigg)\, ,
\een
where $b(q,r,s)$ are integers and
\be \label{ecvalue}
   C = - (i\sqrt N)^{-k-2}\, .
\ee

\end{document}